\documentclass[11pt,twoside,nohyper]{JHEP2} 
\usepackage{epsfig}
\usepackage{amssymb}
\usepackage{latexsym}
\usepackage{times}
\usepackage{slashed,cite}
\usepackage{upgreek}
\usepackage{rotating}
\usepackage{amsmath}
\usepackage{amsfonts}
\usepackage{amsbsy}
\usepackage{amscd}
\usepackage{bbm}

\newcommand{\arxiv}[1]{{\tt arXiv:#1}}

\newcommand{\Eref}[1]{Eq.~(\ref{#1})}
\newcommand{\Sref}[1]{Sec.~\ref{#1}}
\newcommand{\Fref}[1]{Fig.~\ref{#1}}
\newcommand{\Tref}[1]{Table~\ref{#1}}
\newcommand{\cref}[1]{Ref.~\cite{#1}}
\newcommand{\crefs}[1]{Refs.~\cite{#1}}

\newcommand{\bal}{\begin{align}}
\newcommand{\eal}{\end{align}}
\newcommand{\beqs}{\begin{subequations}}
\newcommand{\eeqs}{\end{subequations}}
\newcommand{\ec}{\end{center}}
\newcommand{\bec}{\begin{center}}

\newcommand{\eem}{\end{matrix}}
\newcommand{\bem}{\begin{matrix}}
\newcommand{\eeq}{\end{equation}}
\newcommand{\beq}{\begin{equation}}
\newcommand{\ba}{\begin{array}}
\newcommand{\ea}{\end{array}}
\newcommand{\bea}{\begin{eqnarray}}
\newcommand{\eea}{\end{eqnarray}}
\newcommand{\baq}{\begin{eqnarray}}
\newcommand{\eaq}{\end{eqnarray}}

\newcommand\eqs[2]{Eqs.~(\ref{#1}) and (\ref{#2})}

\newcommand{\ftn}{\footnotesize}

\newcommand{\ssz}{\scriptsize}

\newcommand{\GeV}{{\mbox{\rm GeV}}}

\newcommand{\sFref}[2]{Fig.~\ref{#1}-{\ftn\sf ({#2})}}
\newcommand{\sFsref}[2]{Figs.~\ref{#1}-{\ftn\sf ({#2})}}
\newcommand{\sEref}[2]{Eq.~(\ref{#1}{\ftn\sf {#2}})}

\newcommand{\etal}{{\it et al.\/}}

\def\to{\rightarrow}

\def\llgm{\left\lgroup}
\def\rrgm{\right\rgroup}
\def\lf{\left(}
\def\rg{\right)}
\newcommand\vev[1]{\langle {#1} \rangle}

\newcommand{\Vhi}{\ensuremath{\widehat V_{\rm HI}}}
\newcommand{\dV}{\ensuremath{\Delta\widehat V_{\rm HI}}}
\newcommand{\Hhi}{\ensuremath{\widehat H_{\rm HI}}}
\newcommand{\Ohi}{\ensuremath{\Omega}}
\newcommand{\Omg}{\ensuremath{\Omega}}
\newcommand{\Khi}{\ensuremath{K}}
\newcommand{\Whi}{\ensuremath{W}}
\newcommand{\Vhio}{\ensuremath{\widehat V_{\rm HI0}}}

\newcommand{\mP}{\ensuremath{m_{\rm P}}}
\newcommand{\Mpq}{\ensuremath{M}}

\newcommand{\Mgut}{\ensuremath{M_{\rm GUT}}}
\newcommand{\Qef}{\ensuremath{\Lambda_{\rm UV}}}

\newcommand{\Gbl}{\ensuremath{G_{B-L}}}
\newcommand{\Gsm}{\ensuremath{G_{\rm SM}}}

\def\openone{\leavevmode\hbox{\small1\kern-3.8pt\normalsize1}}

\newcommand{\arcsinh}{\ensuremath{{\rm arcsinh}}}

\newcommand{\ca}{\ensuremath{c_{\cal R}}}
\newcommand{\fr}{\ensuremath{f_{\cal R}}}
\newcommand{\fns}{\ensuremath{f_{0\star}}}
\newcommand{\fnns}{\ensuremath{f_{n\star}}}
\newcommand{\Fs}{\ensuremath{F_{-}}}
\newcommand{\Fvs}{\ensuremath{F_{+}}}
\newcommand{\Fk}{\ensuremath{F_{S}}}
\newcommand{\re}{\ensuremath{e_n}}
\newcommand{\kx}{\ensuremath{k_S}}
\newcommand{\cm}{\ensuremath{c_{-}}}
\newcommand{\cp}{\ensuremath{c_{+}}}

\newcommand{\msn}{\ensuremath{\what m_{\rm \dph}}}

\newcommand{\ns}{\ensuremath{n_{\rm s}}}
\newcommand{\as}{\ensuremath{a_{\rm s}}}
\newcommand{\As}{\ensuremath{A_{\rm s}}}
\newcommand{\rcc}{\ensuremath{\mathcal{R}}}
\newcommand{\rce}{\ensuremath{\widehat{\mathcal{R}}}}
\newcommand{\Ve}{\ensuremath{\widehat{V}}}

\newcommand{\Ne}{\ensuremath{{\what N}}}
\newcommand{\Ns}{\ensuremath{\what N_{\star}}}
\newcommand{\sni}{\ensuremath{\nu^c_i}}

\newcommand{\rw}{\ensuremath{r_{0.002}}}
\newcommand{\rs}{\ensuremath{r_{\pm}}}

\newcommand{\Na}{\ensuremath{{N_1}}}
\newcommand{\Nb}{\ensuremath{{N_2}}}
\newcommand{\Ka}{\ensuremath{{K_1}}}
\newcommand{\Kb}{\ensuremath{{K_2}}}

\newcommand{\phc}{\ensuremath{\Phi}}
\newcommand{\phcb}{\ensuremath{\bar\Phi}}
\newcommand{\dphi}{\ensuremath{\what{\delta\phi}}}
\newcommand{\dph}{\ensuremath{\delta\phi}}

\newcommand{\what}{\ensuremath{\widehat}}

\def\aal{{\bar\alpha}}
\def\bbet{{\bar\beta}}
\def\al{{\alpha}}

\def\bt{{\beta}}

\def\th{{\theta}}
\def\thp{{\theta_+}}
\def\thm{{\theta_-}}
\def\thb{{\bar\theta}}

\def\thn{{\theta_{\Phi}}}

\newcommand{\Trh}{\ensuremath{T_{\rm rh}}}
\newcommand{\sg}{\ensuremath{\phi}}

\newcommand{\sgx}{\ensuremath{\phi_\star}}
\newcommand{\sgf}{\ensuremath{\phi_{\rm f}}}

\newcommand{\ld}{\ensuremath{\lambda}}
\newcommand{\ldu}{\ensuremath{\uplambda}}
\newcommand{\Ld}{\ensuremath{\Lambda}}
\newcommand{\kp}{\ensuremath{\kappa}}
\newcommand{\se}{\ensuremath{\widehat \phi}}
\newcommand{\sex}{\ensuremath{\widehat{\phi}_\star}}

\newcommand{\geu}{\ensuremath{\widehat g}}
\newcommand{\eph}{\ensuremath{\widehat \epsilon}}
\newcommand{\ith}{\ensuremath{\widehat \eta}}

\def\trns{transplanckian}
\def\Kap{K\"{a}hler potential}

\def\sub{subplanckian}

\newcommand{\diag}{\ensuremath{{\sf diag}}}

\newcommand{\tr}{{\mbox{\sf\ssz T}}}

\newcommand\mtt[4]{\mbox{
$\llgm\bem #1 &#2 \cr #3& #4\eem\rrgm$}}

\def\bcp{B{\sc\small icep2}/{\it Keck Array}}
\newcommand{\bicep}{{B{\scshape icep}2}}
\newcommand{\plk}{{\it Planck}}

\title{Shift Symmetry and Higgs Inflation in Supergravity with
Observable Gravitational Waves}

\author{G. Lazarides$^{\sf(1)}$ and C. Pallis$^{\sf(2)}$ \\
$^{\sf(1)}$ School of Electrical \& Computer Engineering, Faculty
of Engineering, \\  Aristotle University of Thessaloniki,
Thessaloniki, GR-54124 Greece; \\ E-mail:
\email{lazaride@eng.auth.gr}\\
$^{\sf(2)}$ Departament de F\'isica Te\`orica and IFIC, \\
Universitat de Val\`encia-CSIC, E-46100 Burjassot, SPAIN; \\
E-mail: \email{cpallis@ific.uv.es}}

\abstract{We demonstrate how to realize within supergravity a
novel chaotic-type inflationary scenario driven by the radial
parts of a conjugate pair of Higgs superfields causing the
spontaneous breaking of a grand unified gauge symmetry at a scale
assuming the value of the supersymmetric grand unification scale.
The superpotential is uniquely determined at the renormalizable
level by the gauge symmetry and a continuous $R$ symmetry. We
select two types of \Kap s, which respect these symmetries as well
as an approximate shift symmetry. In particular, they include in a
logarithm a dominant shift-symmetric term proportional to a
parameter $\cm$ together with a small term violating this symmetry
and characterized by a parameter $\cp$. In both cases, imposing a
lower bound on $\cm$, inflation can be attained with \sub\ values
of the original inflaton, while the corresponding effective theory
respects perturbative unitarity for $\rs=\cp/\cm\leq1$. These
inflationary models do not lead to overproduction of cosmic
defects, are largely independent of the one-loop radiative
corrections and accommodate, for natural values of $\rs$,
observable gravitational waves consistently with all the current
observational data. The inflaton mass is mostly confined in the
range $(3.7-8.1)\times 10^{10}~\GeV$.

\\ \\
{\ftn \sf Keywords: Cosmology of Theories Beyond the Standard
Model, Supergravity Models};\\
{\ftn \sf PACS codes: 98.80.Cq, 11.30.Qc, 12.60.Jv, 04.65.+e}
\\\\{\sl\bfseries Published in} {\sl J. High Energy Phys.  \textbf{11},
114 (2015)}}

\begin{document}




\section{Introduction}\label{intro} %

The announcement of the recent joint analysis of the \bcp\ and
\plk\ data \cite{gws,plcp} confirms earlier attempts \cite{gws2,
kdust} which stressed the impact of the dust foreground on the
observations on the B-mode in the polarization of the cosmic
microwave background radiation at large angular scales. As a
consequence, the predicted tensor-to-scalar ratio $r$ in
inflationary models must be smaller than the one initially claimed
in \cref{gws1}. However, the present data not only leave open the
possibility for a sizable value of $r$, but also seem to favor
values of $r$ of order $0.01$. Indeed, it has been reported that
\beq \label{rgw} r=0.048^{+0.035}_{-0.032} \eeq at 68$\%$
\emph{confidence level} ({\sf\ftn c.l.}). This fact motivates us
to explore the question whether realistic \emph{supersymmetric}
({\sf\ftn SUSY}) inflation models can accommodate such values of
$r$ -- for similar attempts see \crefs{aroest, rEllis, nIG, nMCI,
nMkin}.

One elegant SUSY model which can nicely combine inflation with the
Higgs mechanism of the symmetry breaking is the model of
\emph{Higgs inflation} ({\sf \ftn HI}). This is an inflationary
model of the chaotic type, where a Higgs field plays the role of
inflaton before its trapping in the vacuum. It has been shown that
HI in the framework of \emph{supergravity} ({\sf \ftn SUGRA}) can
be implemented by imposing \cite{shift,shiftHI, takahashi, bauman,
khalil} a convenient shift symmetry on the \Kap\ or invoking
\cite{linde1,nmH,jones2} a logarithmic \Kap\ with a real
subdominant kinetic part and a dominant holomorphic (and
anti-holomorphic) part, which plays the role of a non-minimal
coupling to the Ricci scalar curvature \cite{old, sm1}. Inspired
by these efforts, we present here a `hybrid' scenario, where a
conjugate pair of Higgs superfields is involved in the logarithmic
part of the \Kap\ $K$, which respects a shift symmetry with a tiny
violation -- cf. \crefs{linde2,roest, nMkin}. The
shift-symmetry-preserving part of $K$ influences the amplitude of
the canonically normalized inflaton, which becomes much larger
than the original inflaton field appearing in the superpotential
and the \Kap\ -- cf. \cref{nIG}. Therefore, HI can be implemented
even with \sub\ values of the original inflaton field, keeping
corrections from higher order terms harmless. On the other hand,
the resulting inflationary potential does not depend directly on
the strength of the shift-symmetry-preserving part of $K$ and,
thus, is not flattened drastically as in the original scenario of
non-minimal HI \cite{jones2, linde1,nmH}, but just adequately by
the shift-symmetry-violating part of $K$, as in the recently
proposed models of kinetically modified non-minimal inflation
\cite{nMkin}. Moreover, invoking deviations from the prefactors
$-3$ or $-2$ of the logarithms in the proposed K\"{a}hler
potentials, we succeed  to enhance the resulting values of $r$ --
cf. \crefs{aroest, nIG, nMCI}. We also analyze the impact of the
one-loop \emph{radiative corrections} ({\sf\ftn RCs}) \cite{cw} on
our results and find that these can be kept under control provided
that the relevant renormalization scale is
conveniently chosen \cite{Qenq}. We, finally, show that the
\emph{ultraviolet} ({\sf\ftn UV}) cut-off scale
\cite{cutoff,cutof,riotto} in these models coincides with the
Planck scale and so concerns regarding their naturalness can be
safely eluded.

We exemplify our proposal in the context of a \emph{grand unified
theory} ({\sf\ftn GUT}) model based on the gauge group $G_{B-L}=
G_{\rm SM}\times U(1)_{B-L}$ -- where ${G_{\rm SM}}= SU(3)_{\rm c}
\times SU(2)_{\rm L}\times U(1)_{Y}$ is the gauge group of the
standard model and $B$ and $L$ denote the
baryon and lepton number, respectively. Actually, this is a minimal
extension of the \emph{minimal supersymmetric standard model}
({\sf\ftn MSSM}) which is obtained by promoting the already
existing $U(1)_{B-L}$ global symmetry to a local one. As a
consequence, the presence of right-handed neutrinos $\sni$ is
necessary in order to cancel the $B-L$ gauge anomaly. The Higgs
fields which cause the spontaneous breaking of the $\Gbl$ symmetry
to ${G_{\rm SM}}$ can naturally play the role of inflaton. This
breaking provides large Majorana masses to the $\sni$'s, which
then generate the tiny neutrino masses via the seesaw mechanism.
Furthermore, the out-of-equilibrium decay of the $\sni$'s provides
us with a robust baryogenesis scenario via non-thermal
leptogenesis \cite{ntlepto}.

It is worth emphasizing that $U(1)_{B-L}$ is already spontaneously
broken during HI through the non-zero values acquired by the
relevant Higgs fields. Consequently, HI is not followed by the
production of cosmic strings and, therefore, no extra restrictions
\cite{strings} on the model parameters have to be imposed, in
contrast to the case of the standard \emph{F-term hybrid inflation}
({\sf\ftn FHI}) \cite{susyhybrid} models, which share the
same superpotential with our models. In the standard FHI
models, the GUT gauge symmetry is unbroken during inflation
since the Higgs superfields are confined to zero and the
inflaton is a gauge singlet. The spontaneous breaking of the
GUT gauge symmetry takes place at the end of FHI, where the
Higgs fields acquire non-zero values. Topological defects
are, thus, copiously formed if they are predicted by the
symmetry breaking. In our present scheme, this same gauge
singlet superfield is stabilized at zero during and after HI.
We consider two possible embeddings of the gauge singlet
superfield in $K$ with its kinetic terms included or not
included in the logarithm together with those of the
inflaton.

The superpotential and  \Kap s of our models are presented in
Sec.~\ref{fhim}. In Sec.~\ref{hi}, we describe the inflationary
potential at tree level and after including the one-loop RCs,
whereas, in \Sref{fhi}, we derive the inflationary observables and
confront them with observations. We then provide an analysis of
the UV behavior of these models in Sec.~\ref{fhi3}. Our
conclusions are summarized in Sec.~\ref{con}. Throughout the
paper, we use units where the reduced Planck scale $\mP = 2.433
\times 10^{18}~\GeV$ is set equal to unity, subscripts of the type
$,\chi$ denote derivation \emph{with respect to} ({\small\sf
w.r.t.}) the field $\chi$ (e.g. $F_{,\chi\chi}=\partial^2F/
\partial\chi^2$), and charge conjugation is denoted by a star.

\section{Modeling Shift Symmetry for Higgs Inflation}\label{fhim}

We will now explain how a conveniently modified shift symmetry
can be used in order to implement HI based on the F-term SUGRA
potential. The structure of the superpotential is presented in
\Sref{fhim1}, whereas the relevant \Kap\ is given in \Sref{fhim2}.
Finally, in \Sref{fhim3}, we derive the corresponding frame
function.

\subsection{The superpotential} \label{fhim1}

\begin{floatingtable}[r]
\begin{tabular}{|l||lll|}\hline
{\sc Superfields}&$S$&$\Phi$&$\bar\Phi$\\\hline\hline
$U(1)_{B-L}$&$0$&$1$&$-1$\\\hline
$U(1)_{R}$ &$1$&$0$&$0$\\\hline
\end{tabular}
\caption {\sl Charge assignments of the superfields.}\label{ch}
\end{floatingtable}

We focus on a minimal extension of the MSSM based on the gauge group
$G_{B-L}$, which can be broken down to $G_{\rm SM}$ at a scale
close to the SUSY GUT scale $M_{\rm GUT}$ through the vacuum
expectation values acquired by a conjugate pair
of left-handed Higgs superfields $\Phi$ and $\bar\Phi$ charged
oppositely under $U(1)_{B-L}$ -- see \Tref{ch}. The part of the
superpotential $W$ which is relevant for inflation is given by
\cite{susyhybrid}
\beq
W=\ld S\lf\bar\Phi\Phi-\Mpq^2/4\rg\label{Win},
\eeq
where $S$ is a gauge singlet superfield, $\ld$ a dimensionless
parameter, and $M$ a mass scale of order $M_{\rm GUT}$. This
superpotential is the most general renormalizable superpotential
which respects an $R$ symmetry $U(1)_{R}$ -- see \Tref{ch} -- in
addition to the aforementioned $\Gbl$. The $R$ symmetry
guarantees the linearity of the superpotential w.r.t. the gauge
singlet superfield $S$. This fact is helpful both for the
realization of HI and the determination of the SUSY vacuum.

To verify that $W$ leads to the breaking of $G_{B-L}$ down to
$\Gsm$, we minimize the SUSY limit $V_{\rm SUSY}$ of the SUGRA
scalar potential derived from the superpotential in
Eq.~(\ref{Win}) and the common SUSY limit of the K\"{a}hler
potentials in Eqs.~(\ref{K1}) and (\ref{K2}) -- see below.
The potential $V_{\rm SUSY}$, which includes contributions from
F- and D-terms, turns out to be
\beqs
\beq V_{\rm SUSY}= \ld^2\left|\phcb\phc-\frac{M^2}4\right|^2
+\frac{\ld^2}{\cm(1-N\rs)}|S|^2\lf|\phcb|^2+|\phc|^2\rg+
\frac{g^2}2\cm^2(1-N\rs)^2\lf|\phcb|^2-|\phc|^2\rg^2\,,
\label{VF}\eeq
where the complex scalar components of the various superfields are
denoted by the same superfield symbol, $g$ is the unified gauge
coupling constant, and the remaining parameters ($N,\cm,\rs$) are
defined in Secs.~\ref{fhim2} and \ref{fhim3}. From the last
equation, we find that the SUSY vacuum lies along the D-flat
direction $|\phcb|=|\phc|$ with
\beq \vev{S}=0 \>\>\>\mbox{and}\>\>\>
|\vev{\Phi}|=|\vev{\bar\Phi}|=\Mpq/2.\label{vevs} \eeq\eeqs
Although $\vev{\Phi}$ and $\vev{\bar\Phi}$ break spontaneously
$U(1)_{B-L}$, no cosmic strings  are produced at the end of
inflation, since this symmetry is already broken during HI.
Needless to say that contributions from the soft SUSY breaking
terms can be safely neglected since the corresponding mass
scale is much smaller than $M$. Let us emphasize, however,
that soft SUSY breaking effects break $U(1)_R$ explicitly to
a discrete subgroup. Usually, the combination of the latter
with the $\mathbb{Z}_2^{\rm f}$ fermion parity yields
\cite{chios} the well-known $R$-parity of MSSM, which
guarantees the stability of the lightest SUSY particle and,
therefore, provides a well-motivated cold dark matter
candidate.

\subsection{The K\"{a}hler Potential} \label{fhim2}

The superpotential $W$ in \Eref{Win} could give rise to HI driven
by the real field $\sg$ defined in the standard parametrization
\beq\label{hpar} \Phi=\frac{\sg}{\sqrt{2}}
e^{i\th}\cos\thn,\>\>\>\bar\Phi=\frac{\sg}{\sqrt{2}}
e^{i\thb}\sin\thn\>\>\>\mbox{with}\>\>\>0\leq\thn\leq\frac{\pi}{2}
\>\>\>\mbox{and}\>\>\>S=
\frac{s +i\bar s}{\sqrt{2}}\eeq
provided that we confine ourselves to the field configuration
\beq\label{inftr}
s=\bar s=\th=\thb=0\>\>\>\mbox{and}\>\>\>\thn={\pi/4}.\eeq
Note that the last equality ensures the D-flatness of the potential.
Indeed, along this trajectory, $V_{\rm SUSY}$ in \Eref{VF} reduces
to the well-known F-term potential which is quartic w.r.t. $\sg$.
This construction, though, can become meaningful only if it can be
successfully embedded in SUGRA, due to the \trns\ values of $\phi$
which may be needed -- see below.

To this end and following similar works \cite{shift,shiftHI,khalil},
we require that the \Kap\ is consistent with the shift symmetry
\beqs\beq \Phi\ \to\ \Phi+C,~\bar\Phi\ \to\ \bar\Phi+ C^*,~S \to\
S\label{shift}\eeq
with $C$ being any complex number. Under this symmetry, the
real quantities
\beq \Fs=\left|\Phi-\bar\Phi^*\right|^2
~~~\mbox{and}~~~\Fk={|S|^2}-{\kx}{|S|^4}\label{Fs}\eeq\eeqs
are invariant and can, thus, be used in the construction of the
K\"{a}hler potential. The last term in the right-hand side
of the equation for \Fk\ with $\kx\sim1$ is included in
order to ensure that the mass squared of $S$ during HI is large
and positive -- see \Sref{hi2}. If one combines the superpotential in
\Eref{Win} with a canonical-like \cite{shiftHI} or a logarithmic
\Kap\ involving $\Fs$ and $\Fk$, one can show that HI driven by the
simplest quartic potential with \trns\ values of $\phi$ can be
attained. Namely, in \cref{shiftHI}, a symmetry similar to the
one in \Eref{shift} is conveniently applied in the case of the
$SU(2)_{\rm L}$ doublets of MSSM. This model, though, is by now
ruled out \cite{plcp} due to the relatively low value of $\ns$
($\ns\simeq0.947$) and the high value of $r$ ($r\simeq0.28$)
that it predicts -- see e.g. \crefs{talk,plcp}.

We are forced, therefore, to allow a tiny violation of the
shift symmetry in \Eref{shift} including at the level of the
quadratic terms in the K\"{a}hler potential a subdominant
term of the form
\beq \label{Fvs}\Fvs=\left|\Phi+\bar\Phi^*\right|^2,\eeq
which remains invariant under the transformation
\beq \Phi\ \to\ \Phi+ C,~\bar\Phi\ \to\ \bar\Phi-
C^*,~S\ \to\ S
\label{shift1}\eeq
coinciding with the one in \Eref{shift} only for $C=0$. Both $\Fs$
and $\Fvs$ respect the symmetries imposed on $W$ and generate
kinetic mixing between $\phc$ and $\phcb$ with non-vanishing
eigenvalues. However, positive eigenvalues of the kinetic mixing
of $\Phi$ and $\bar\Phi$ for a logarithmic \Kap\ are provided by
$\Fs$, which has, thus, to play a prominent role -- see \Sref{hi1}.
The quantity $\Fs$, contrary to $\Fvs$, vanishes along the
trajectory in \Eref{inftr} and it is expected to contribute
only to the normalization of the inflaton field, whereas $\Fvs$
is expected to have an impact on both the normalization of the
inflaton and the inflationary potential.

Including the terms in \eqs{Fs}{Fvs} in a canonical-like \Kap, we
obtain a model which can become just marginally compatible with the
data as mentioned in \cref{shiftHI}. Here, we adopt two other
alternatives, i.e. a purely logarithmic \Kap
\beq
K_1=-\Na\ln\lf1-\frac\cm\Na\Fs+\cp\Fvs-\frac1\Na\Fk+\frac{k_{-}}
{\Na}\Fs^2+\frac{k_{S-}}{\Na}\Fs|S|^2\rg\,,\label{K1}\eeq
or a `hybrid' \Kap\
\beq
K_2=-\Nb\ln\lf1-\frac{\cm}{\Nb}\Fs+\cp\Fvs\rg+\Fk\,\label{K2}
\eeq
with one logarithmic term for $\Phi$ and $\bar\Phi$ and one
canonical-like kinetic term for $S$.
In both cases, positivity of the kinetic energy requires $\Na>0$
and $\Nb>0$. We also introduced two dimensionless coupling constants
$\cm$ and $\cp$ with a clear hierarchy $\cm\gg\cp$ so that the
shift symmetry in \Eref{shift} is the dominant symmetry of the \Kap\
compared to that in \Eref{shift1}. It is worth mentioning that our
models are completely natural in the 't Hooft sense because, in the
limit $\cp\to0$ and $\ld\to0$, the shift symmetry in \Eref{shift}
becomes exact and a $U(1)$ symmetry under which $S \to\ e^{i\alpha} S$
($\alpha$ is a real number) appears.

Note, finally, that the scenario of non-minimal HI investigated in
\crefs{linde1,nmH} can be recovered by doing the substitutions
\beq\label{cR} \cm=1+N\ca~~\mbox{and}~~\cp=\ca~~\eeq
with $N=\Na=3$ in \Eref{K1} or $N=\Nb=2$ in \Eref{K2} -- see
below. The symmetries of our model, though, prohibit the existence
in the \Kap\ of the terms $|S|^2\lf k_{S\Phi} |\phc|^2+
k_{S\bar\Phi}|\phcb|^2\rg$ which, generally, violate \cite{talk}
D-flatness. In this case, the restoration of the D-flatness would
require the equality of $k_{S\Phi}$ and $k_{S\bar\Phi}$, which
signals an ugly tuning of the parameters.
\subsection{The Frame Function}\label{fhim3}

The interpretation of the \Kap s in \eqs{K1}{K2} can be given
in the `physical' \emph{Jordan frame} ({\ftn\sf JF}). To this end,
we derive the JF action for the scalar fields $z^\al=S,\phc,\phcb$.
We start with the corresponding \emph{Einstein frame}
({\small\sf EF}) action within SUGRA \cite{linde1,talk}, which
can be written as
\beqs\beq\label{Saction1} {\sf S}=\int d^4x \sqrt{-\what{
\mathfrak{g}}}\lf-\frac{1}{2} \rce +K_{\al\bbet}\geu^{\mu\nu}
D_\mu z^\al D_\nu z^{*\bbet}-\Ve\rg, \eeq
where $\widehat{\mathfrak{g}}$ is the determinant of the EF metric
$\geu_{\mu\nu}$, $\rce$ is the EF Ricci scalar curvature, $D_\mu$
is the gauge covariant derivative, and $\Ve$ is the (tree-level) EF
SUGRA scalar potential given by
\beq \Ve=\Ve_{\rm F}+ \Ve_{\rm D}\>\>\>\mbox{with}\>\>\> \Ve_{\rm
F}=e^{\Khi}\left(K^{\al\bbet}{\rm F}_\al {\rm F}^*_\bbet-3\vert
W\vert^2\right) \>\>\>\mbox{and}\>\>\>\Ve_{\rm D}= {1\over2}g^2
\sum_a {\rm D}_a {\rm D}_a. \label{Vsugra} \eeq
Here, a trivial gauge kinetic function is adopted and the
summation is applied over the generators $T_a$ of the considered
gauge group. Also, we use the shorthand notation
\beq \label{Kinv} K^{\bbet\al}K_{\al\bar
\gamma}=\delta^\bbet_{\bar \gamma},\>\>{\rm F}_\al=W_{,z^\al}
+K_{,z^\al}W,~~\mbox{and}~~{\rm D}_a=z^\al\lf T_a\rg_\al^\bt
K_\bt~~\mbox{with}~~K_{\al}={\Khi_{,z^\al}},~~K_{\al\bbet}=
{\Khi_{,z^\al z^{*\bbet}}}.\eeq\eeqs

If we perform a conformal transformation \cite{linde1, talk}
defining the JF metric $g_{\mu\nu}$ through the relation
\beqs\beq \label{weyl}
\geu_{\mu\nu}=-\frac{\Omega}{N}g_{\mu\nu},\>\>\>\mbox{we
obtain}\>\>\>\left\{\bem
\sqrt{-\what{ \mathfrak{g}}}={\Omega^2\over
N^2}\sqrt{-\mathfrak{g}},\>\>\>
\geu^{\mu\nu}=-{N\over\Omega}g^{\mu\nu}, \hfill \cr
\mbox{and}\>\>\>\rce=-{N\over\Omega}\left(\rcc-\Box\ln \Omega+3g^{\mu\nu}
\partial_\mu \Omega\partial_\nu \Omega/2\Omega^2\right). \hfill
\cr\eem
\right.\eeq
Here $\Omega$ is the frame function, $\mathfrak{g}$ is the
determinant of $g_{\mu\nu}$, $\rcc$ is the JF Ricci scalar
curvature and $N$ is a dimensionless parameter which quantifies
the deviation from the standard set-up \cite{linde1}. Upon
substitution of \Eref{weyl} into \Eref{Saction1}, we end up with
the following action in the JF
\beq {\sf S}=\int d^4x \sqrt{-\mathfrak{g}}\lf\frac{\Omega}{2N}
\rcc+\frac{3}{4N\Omega}D_\mu\Omega D^\mu\Omega -\frac{1}{N}\Omega
K_{\al{\bbet}}D_\mu z^\al D^\mu z^{*\bbet}-V
\rg\>\>\>\mbox{with}\>\>\>V=\frac{\Omg^2}{N^2}\Ve\,.
\label{action2}\eeq\eeqs
If, in addition, we connect $\Omega$ to $K$ through the following
relation
\beqs\beq-\Omega/N =e^{-K/N }\>\Rightarrow\>K=-N
\ln\lf-\Omega/N\rg\label{Omg1}\eeq
and take into account the definition \cite{linde1} of the purely
bosonic part of the on-shell value of the auxiliary field
\beq {\cal A}_\mu =i\lf K_\al D_\mu z^\al-K_\aal D_\mu
z^{*\aal}\rg/6, \label{Acal1}\eeq
we arrive at the following action
\beq {\sf S}=\int d^4x \sqrt{-\mathfrak{g}}\lf\frac{
\Omega}{2N}\rcc+\lf\Omega_{\al{\bbet}}+\frac{3-N}{N}
\frac{\Omega_{\al}\Omega_{\bbet}}{\Omega}\rg D_\mu z^\al D^\mu
z^{*\bbet}- \frac{27}{N^3}\Omega{\cal A}_\mu{\cal A}^\mu-V \rg,
\label{Sfinal}\eeq
where ${\cal A}_\mu$ in \Eref{Acal1} takes the form
\beq {\cal A}_\mu =-iN \lf \Omega_\al D_\mu z^\al-\Omega_\aal
D_\mu z^{*\aal}\rg/6\Omega\,\label{Acal}\eeq\eeqs
and the shorthand notation $\Omega_\al=\Omega_{,\Phi^\al}$
and $\Omega_\aal=\Omega_{,\Phi^{*\aal}}$ is used. From
\Eref{Omg1}, we can find the corresponding frame function
during HI as follows
\beq \fr=-\left.{\Ohi\over
N}\right|_{\mbox{\Eref{inftr}}}=\left\{\bem
%
(1+\cp\sg^2)^{\Na/N}\hfill&\mbox{for $K=K_1$,}\hfill\cr
(1+\cp\sg^2)^{\Nb/N}\hfill \>\>\>&\mbox{for $K=K_2$,}\hfill\cr
\eem
\right.\, \label{minK}\eeq
where we took into account the fact that $\Fs=\Fk=0$ along the
direction in \Eref{inftr}. Eqs.~(\ref{Sfinal}) and (\ref{minK})
reveal that $\fr$
represents the non-minimal coupling to gravity. Note that this
function is independent of $\cm$, which is to be large for HI
with $\sg<1$ -- see below. Selecting
\beq \label{Nab} N=\Na~~\mbox{or}~~
N=\Nb~~\mbox{for}~~K=\Ka~~\mbox{or}~~ K=\Kb\eeq
respectively, we can obtain the standard quadratic non-minimal
coupling function. As for the conventional case \cite{linde1, nmH}
with one logarithm and $N=3$, when the dynamics of the fields
$z^\al$ is dominated only by the real moduli $|z^\al|$ or when
$z^\al=0$ for $\al\neq1$ \cite{linde1}, we obtain
${\cal A}_\mu=0$ in \Eref{Acal}. The only difference w.r.t.
the aforementioned conventional case is that now the scalar fields
$z^\al$ have non-canonical kinetic terms in the JF due to the term
proportional to $\Omg_\al\Omg_\bbet\neq\delta_{\al\bbet}$. This
fact does not cause any problem, since the canonical normalization
of the inflaton retains its strong dependence on $\cm$ through
$\Omega$, whereas the non-inflaton fields become heavy enough
during inflation and so they do not affect the dynamics -- see
\Sref{hi1}. Furthermore, for $M\ll\mP$, the conventional Einstein
gravity at the SUSY vacuum -- in \Eref{vevs} -- is recovered since
\beq -\vev{\Ohi}/N\simeq1. \label{ig}\eeq
Given that the analysis of inflation in both the EF and JF yields
equivalent results \cite{old}, we carry out the derivation of the
inflationary observables exclusively in the EF -- see
Secs.~\ref{hi1} and \ref{hi2}.

\section{The Inflationary Set-up}\label{hi}

In this section, we outline the salient features of our
inflationary scenario (\Sref{hi1}) and then present the
one-loop corrected inflationary potential in Sec.~\ref{hi2}.

\subsection{The Tree-level Inflationary Potential}\label{hi1}

The linearity of $\Whi$ w.r.t. $S$ allows us to isolate easily the
non-vanishing contribution of the inflaton to $\Ve_{\rm F}$ on the
inflationary path and avoid the runaway behavior which may be caused
by the term  $-3|W|^2\exp{K}$. Indeed, inserting \eqs{Win}{K1} or
(\ref{K2}) into \Eref{Vsugra}, we find that the only surviving
contribution to $\Ve$ on the path in \Eref{inftr} is
\beq \Vhio= e^{K}K^{SS^*}\,
|W_{,S}|^2=\frac{\ld^2(\sg^2-M^2)^2}{16}\times\left\{\bem
%
\fr^{-\Na+1}\hfill&\mbox{for $K=K_1$}\hfill\cr
\fr^{-\Nb}\hfill \>\>\>&\mbox{for $K=K_2$.}\hfill\cr \eem
\right. \label{1Vhio}\eeq
Here we took into account \eqs{minK}{Nab} and the fact that
$e^{K}=\fr^{-N}$ and $K^{SS^*}=\fr$ or $K^{SS^*}=1$ for $K=K_1$ or
$K_2$ respectively -- note that $K_{Sz^\al}=0$ for both cases and
$\al=2$ or $3$. Introducing a new variable $n$ related to the
exponents of $\fr$ in Eq.~(\ref{1Vhio}), we can cast $\Vhio$ in
the same form for both the $K$'s in \eqs{K1}{K2}. Indeed, $\Vhio$
can be rewritten as
\beq \Vhio=\frac{\ld^2(\sg^2-M^2)^2}{16\fr^{2(1+n)}},
\>\>\mbox{where}\>\> \left\{\bem
%
\Na-1=2(1+n)\hfill\cr
\Nb=2(1+n)\hfill\cr \eem
\right. \>\>\mbox{for}\>\> \left\{\bem
%
K=K_1\hfill\cr
K=K_2.\hfill\cr \eem
\right.\label{2Vhio}\eeq
As anticipated below \Eref{shift1}, $\Vhio$ depends exclusively on
$\cp$ (and not on $\cm$). Given that, during HI, $\sg\gg M$ and
$\cp\sg^2>1$ -- see below --, $\Vhio$ and the corresponding Hubble
parameter $\Hhi$ take the form
\beq
\Vhio\simeq\frac{\ld^2\sg^{-4n}}{16\cp^{2(1+n)}}\,\>\>\mbox{and}\>\>\>
\Hhi={\Vhio^{1/2}\over\sqrt{3}}\simeq{\ld\sg^{-2n}\over4\sqrt{3}
\cp^{1+n}}\cdot
\label{3Vhio}\eeq
As a consequence, we obtain an inflationary plateau for $n=0$ or a
bounded from below chaotic-type inflationary potential for
$n<0$ with $\sg$ in \Eref{hpar} being a natural inflaton
candidate. Note that, thanks to the shift symmetry in \Eref{shift},
no mixing term proportional to $k_{S-}$ arises in $\Vhio$ in sharp
contrast with the models of \crefs{nIG, nMCI}, where a similar term
($\propto k_{S\Phi}$) plays a crucial role in achieving large values
of $r$ in a manner compatible with all observations -- see \Sref{fhi2}.

To specify further our inflationary scenario, we have to determine
the EF canonically normalized fields involved. Note that, along
the configuration in \Eref{inftr}, the K\"{ahler metric}
$K_{\al\bbet}$ defined in \Eref{Kinv} takes, for both choices of $K$
in \eqs{K1}{K2}, the form
\beq \lf K_{\al\bbet}\rg=\diag\lf M_K,K_{SS^*}\rg,~~\mbox{where}~~
M_K=\frac{1}{\fr^2}\mtt{\kappa}{\bar\kappa}{\bar\kappa}{\kappa}~~
\mbox{with}~~\kappa=\cm\fr-N\cp,~~\bar\kappa={N\cp^2\sg^2},
\label{Sni1} \eeq
and $N$ defined in \Eref{Nab}. Given that $K_{SS^*}=1/\fr$ or
$1$ for $K=K_1$ or $K_2$ respectively, the canonically
normalized components $\what s,\what{\bar{s}}$ of $S$ -- see
\Eref{hpar} -- are defined as follows:
\beq (\what s,\what{\bar{s}})=\sqrt{K_{SS^*}}(s,\bar s). \eeq
The matrix $M_K$ can be diagonalized via a similarity
transformation involving an orthogonal matrix $U_K$ as
follows:
\beq \label{diagMk} U_K M_K U_K^\tr
=\diag\lf\kp_+,\kp_-\rg,\>\>\>\mbox{where}\>\>\>U_K=
{1\over\sqrt{2}}\mtt{1}{1}{-1}{1}\eeq
and the eigenvalues of $M_K$ are found to be
\beq
\label{mK}\kp_+=\frac{\cm(1-N\rs)+\cp\cm(1+N\rs)
\sg^2}{\fr^2}\simeq\frac{\cm}{\fr}\>\>\>
\mbox{and}\>\>\>\kp_-=\frac{\cm(1-N\rs)}{\fr}\simeq
\frac{\cm}{\fr},\eeq
where the approximate results hold for $\rs\ll 1/N$ and positivity
of $\kp_-$ can be assured only if
\beq\label{pos} \rs<1/N~~~\mbox{with}~~~\rs=\cp/\cm,\eeq
i.e. if $\cm>N\cp$, as we anticipated below \Eref{shift1}. This
fact has to be contrasted with the original scenario of
non-minimal HI \cite{linde1,nmH}, where such a constraint is not
necessary -- as can be verified by inserting \Eref{cR} into
\Eref{mK}. In our present cases, the kinetic terms for
$z^\al=\phcb,\phc$ can be brought into the following form
\beqs\beq K_{\al\bbet}\dot z^\al \dot z^{*\bbet}= {\kp_+\over
2}\lf\dot \sg^2+{1\over2}\sg^2\dot\theta^2_+ \rg+{\kp_-\sg^2\over
2}\lf{1\over2}\dot\theta^2_- +\dot\theta^2_\Phi
\rg=\frac12\lf\dot{\widehat \sg}^2+\dot{\widehat
\th}_+^2+\dot{\widehat \th}_-^2+\dot{\widehat \th}_\Phi^2\rg,
\label{Snik}\eeq
where $\th_{\pm}=\lf\bar\th\pm\th\rg/\sqrt{2}$ and the dot denotes
derivation w.r.t. the cosmic time $t$. In the last step, we
introduce the EF canonically normalized fields, which are denoted
by hat and can be obtained as follows:
\beq \label{VJe}
\frac{d\se}{d\sg}=J=\sqrt{\kp_+},\>\>\widehat{\theta}_+
={J\sg\theta_+\over\sqrt{2}},\>\>\widehat{\theta}_-
=\sqrt{\frac{\kp_-}{2}}\sg\theta_-,\>\>\>\mbox{and}\>\>\>\widehat
\theta_\Phi = \sg\sqrt{\kp_-}\lf\theta_\Phi-{\pi\over4}\rg
\cdot\eeq\eeqs
Note, in passing, that the spinors $\psi_S$ and $\psi_{\Phi\pm}$
associated with the superfields $S$ and $\Phi_\pm=(\Phi\pm\bar\Phi)/
\sqrt{2}$ are normalized similarly, i.e.
$\what\psi_{S}=\sqrt{K_{SS^*}}\psi_{S}$ and
$\what\psi_{\Phi\pm}=\sqrt{\kp_\pm}\psi_{\Phi\pm}$.
Integrating the first equation in \Eref{VJe}, we can express the
canonically normalized EF real field $\se$ as follows:
\beq \se=\se_{\rm
c}+\frac1{\sqrt{\rs}}\arcsinh\sqrt{\cp}\sg\label{se1}\eeq
with $\se_{\rm c}$ being a constant of integration, which we take
equal to zero. Note that $\se$ is practically independent of $N$
(and $n$) -- see \Eref{mK}. Solving \Eref{se1} w.r.t. $\sg$, we
can express $\Vhio$ in \Eref{3Vhio} in terms of $\se$ as follows:
\beq \label{Vhin} \Vhio\simeq
\frac{\ld^2}{4\cp^2}\frac{\tanh^4\sqrt{\rs}\se}{\cosh^{4n}
\sqrt{\rs}
\se}\,\cdot\eeq
For $n=0$ and $\ld=\cp=1$, $\Vhio$ coincides with the potential
encountered in the so-called $T$-models \cite{aroest} arising from
the spontaneous breaking of (super)conformal invariance. We
observe that, although $\fr$ in \Eref{minK} and $\Vhio$ in
\Eref{Vhin} are independent of $\cm$, $\se$ depends heavily on
$\cm$ and, therefore, it can be much larger than $\sg$ facilitating
the attainment of HI with \sub\ values of $\sg$. As a consequence,
the initial fields $\Phi$ and $\bar\Phi$ -- see \Eref{hpar} --,
which are closely related with $\sg$, can remain also \sub, as
required for a meaningful approach to SUGRA.

\subsection{Stability and One-loop Radiative Corrections}\label{hi2}

To ensure the validity of our inflationary proposal, we have
to check the stability of the direction in \Eref{inftr} w.r.t. the
fluctuations of the fields which are orthogonal to this direction,
i.e. we have to examine the fulfillment of the following
conditions:
\beqs\beq 
\left.{\partial
\Ve\over\partial\what\chi^\al}\right|_{\mbox{\Eref{inftr}}}=0\>\>\>
\mbox{and}
\>\>\>\what m^2_{
\chi^\al}>0\>\>\>\mbox{with}\>\>\>\chi^\al= \thm,\thp,\thn,s,\bar
s.\label{Vcon} \eeq
Here $\what m^2_{\chi^\al}$ are the eigenvalues of the mass-squared
matrix with elements
\beq \label{wM2}\what
M^2_{\al\bt}=\left.{\partial^2\Ve\over\partial\what\chi^\al
\partial\what\chi^\beta}\right|_{\mbox{\Eref{inftr}}}
\mbox{with}~~\chi^\al=\thm,\thp,\thn,s,\bar s.\eeq\eeqs
Diagonalizing $\what M^2_{\al\bt}$, we construct the scalar
mass-squared spectrum of the theory along the direction in
\Eref{inftr}. In \Tref{tab1}, we present approximate expressions
of the relevant masses squared, which are quite close to the
rather lengthy exact expressions. Note, however, that our
numerical computation uses the exact expressions. In this table,
we also include the mass squared $\what m_\sg^2$ of $\what \phi$
as well as the masses squared of the chiral fermions, the gauge
boson $A_{BL}$, and the gaugino $\lambda_{BL}$ which are used in
our analysis below.

From the formulas displayed in \Tref{tab1}, we can infer that the
stability of the path in \Eref{inftr} is assured since \Eref{Vcon}
is fulfilled. In particular, it is evident that $\kx\gtrsim1$
assists us to achieve $\what m^2_{{s}}>0$ for $K=\Ka$ -- in
accordance with the results for similar models in \crefs{linde1,
nIG}. On the other hand, for $K=\Kb$, $\what m^2_{{s}}>0$ even
with $\kx=0$. However, since there is no observational hint
\cite{plcp} for large non-Gaussianity in the cosmic microwave
background, we should make sure that all the $\what
m^2_{\chi^\al}$'s for the scalar fields in \Tref{tab1} except
$\what m^2_\sg$ are greater than $\Hhi^2$ during the last $50-60$
e-foldings of HI. This guarantees that the observed curvature
perturbation is generated wholly by $\sg$ as assumed in
\Eref{Prob} -- see below. Requiring that $\what m^2_{s}\gg\Hhi^2$
entails the existence of a non-vanishing $\kx$ for $K=\Kb$ too.
Due to the large effective masses that the scalars acquire during
HI, they enter a phase of damped oscillations about zero. As a
consequence, the $\sg$ dependence in their normalization -- see
\Eref{VJe} -- does not affect their dynamics.

\renewcommand{\arraystretch}{1.4}

\begin{table}
\bec
\begin{tabular}{|c||c|c||l|l|}\hline
{\sc Fields}&{\sc Eigenstates}& \multicolumn{3}{|c|}{\sc Masses
Squared}\\\cline{3-5} && {\sc 
} &\multicolumn{1}
{|c|}{$K=K_1$}&\multicolumn{1}{|c|}{$K=K_2$}\\
\hline\hline
3 real scalars&$\se$ &$\what
m^2_\sg$&\multicolumn{2}{|c|}{$\what\eta\Hhi^2$}\\\cline{3-5}
&$\widehat\theta_{+}$&$\widehat m_{\theta+}^2$& $6(1-1/\Na)\Hhi^2$
&$6\Hhi^2$\\
&$\widehat \theta_\Phi$ &$\widehat m_{\theta_\Phi}^2$&
$M^2_{BL}+6\lf1-\frac1\Na\rg\Hhi^2$&
$M^2_{BL}+6\Hhi^2$\\
1 complex scalar&$\widehat s, \widehat{\bar{s}}$ &$ \widehat m_{
s}^2$&$6\lf2\kx\fr-\frac1\Na\rg\Hhi^2$&$12\kx\Hhi^2$\\\hline
1 gauge  boson& $A_{BL}$ &
$ M_{BL}^2$&\multicolumn{2}{|c|}{$g^2\cm(1-N\rs )\sg^2/\fr$}\\
\hline
$4$ Weyl spinors & $\what \psi_\pm ={\what{\psi}_{\Phi+}\pm
\what{\psi}_{S}\over\sqrt{2}}$ & $\what m^2_{ \psi\pm}$&
${6(2+\cp(3-\Na)\sg^2)^2\over\cm^2\sg^2\fr}\Hhi^2$
&${6(2+\cp(2-\Nb)\sg^2)^2\over\cm^2\sg^2\fr}\Hhi^2$
\\\cline{4-5}
&$\ldu_{BL}, \widehat\psi_{\Phi-}$&$M_{BL}^2$&\multicolumn{2}
{|c|}{$g^2\cm(1-N\rs )\sg^2/\fr$}\\
\hline
\end{tabular}
\end{center}
\caption{\sl\ftn The mass-squared spectrum for $K=K_1$ and $K=K_2$
along the inflationary trajectory in \Eref{inftr} for $\phi\ll1$.
$N$ is defined in \Eref{Nab} and
$\widehat\eta$ is given by \Eref{srg} -- see below. To avoid very
lengthy formulas, we neglect terms proportional to $M\ll\sg$.}
\label{tab1}
\end{table}
\renewcommand{\arraystretch}{1.}

Considering SUGRA as an effective theory below $\mP$ allows us to use
the well-known Coleman-Weinberg formula \cite{cw} in order to find the
one-loop corrected inflationary potential
\beqs\beq\Vhi=\Vhio+\dV\>\>\>\mbox{with}\>\>\>\dV=
\frac{1}{64\pi^2}\sum_i(-)^{{\rm
F}_i}\what m_i^4\ln \frac{\what m_i^2}{\Lambda^2}\,,\label{Vhic}
\eeq
where the sum extends over all helicity states $i$ of the fields listed in
\Tref{tab1}, ${\rm F}_i$ is the fermion number and $\what m_i^2$ the mass
squared of the $i$th helicity state, and $\Lambda$ is a renormalization
mass scale. The consistent application of this formula requires that the
$\what m_i^2$'s which enter into the sum are:
\begin{itemize}

\item Positive. As a consequence and following the common
practice \cite{nMCI, nIG, sm1}, we neglect the contribution of
$\what m_\sg^2$ to $\dV$ since this mass squared turns out to
be negative in the largest part of the parameter space of our
models. Let us recall here that \Eref{Vhic} is valid only for
a static configuration and is uniquely defined only in the
extremum points. The proper calculation should use the
time-dependent background -- see e.g. \cref{postma}. For the
non-minimal inflation, this is not done up to now, but we do
not expect to get a very unexpected effect if the full
computation is consistently carried out.

\item Much lighter than a momentum cut-off squared, which is here
considered as large as $\mP^2$. As a consequence, we do not take
into account the contributions from $M_{BL}^2$ and $\what
m_{\theta_\Phi}^2$ to $\dV$ since these masses squared are much
larger than $\mP^2$ in our case given that $\cm\gg1$ -- see
below. This stems from the form of $\kp_-$ in \Eref{mK} and does
not occur in the case of the standard non-minimal HI
\cite{sm1, nmH}, as can be checked by substituting \Eref{cR} into
the expressions for these masses squared in \Tref{tab1}.

\end{itemize}
Having in mind the above subtleties and neglecting contributions
from the gravitational sector of the theory, the one-loop RCs read
\beq  \dV = {1\over64\pi^2}\lf \widehat m_{ \th_+}^4\ln{\widehat
m_{ \th_+}^2\over\Lambda^2} +2 \widehat m_{s}^4\ln{m_{\widehat
s}^2\over\Lambda^2} -4\widehat m_{\psi_{+}}^4\ln{\widehat
m_{\psi_{+}}^2\over\Lambda^2} \rg\,.\label{Vrc}\eeq\eeqs
The renormalization scale $\Lambda$ can be determined by
requiring \cite{Qenq} that $\dV(\sgx)=0$ or $\dV(\sgf)=0$.

Let us, finally, stress here that the non-vanishing value of $\sg$
during HI breaks spontaneously $U(1)_{B-L}$ leading to a Goldstone
boson $\th_-$. This is `eaten' by the gauge boson $A_{BL}$, which
then becomes massive. As a consequence, six degrees of freedom
before the spontaneous breaking (four corresponding to the two
complex scalars $\Phi$ and $\bar \Phi$ and two corresponding to
the massless gauge boson $A_{BL}$ of $U(1)_{B-L}$) are
redistributed as follows: three degrees of freedom are associated
with the real propagating scalars ($\se, \what \th_+$, and $\what
\theta_\Phi$), whereas the residual one degree of freedom
combines together with the two ones of the initially massless
gauge boson $A_{BL}$ to make it massive. From \Tref{tab1}, we can
deduce that the numbers of bosonic (eight) and fermionic (eight)
degrees of freedom are equal, as they should.

\section{Constraining the Parameters of the Models}\label{fhi}

We will now outline the predictions of our inflationary
scenarios in Sec.~\ref{res} and confront them with a number of
criteria introduced in Sec.~\ref{fhi2}.

\subsection{Inflationary Observables -- Constraints} \label{fhi2}

Our inflationary settings can be characterized as successful if
they can be compatible with a number of observational and
theoretical requirements, which are enumerated in the following --
cf. \cref{review}:

\paragraph{4.1.1.} The number of e-foldings
\begin{equation}
\label{Nhi}  \Ns=\int_{\se_{\rm f}}^{\se_\star}\, d\se\:
\frac{\Ve_{\rm HI}}{\Ve_{\rm HI,\se}}= \int_{\sgf}^{\sgx}\,
J^2\frac{\Ve_{\rm HI}}{\Ve_{\rm HI,\sg}}d\sg
\end{equation}
that the pivot scale $k_\star=0.05/{\rm Mpc}$ suffers during HI
has to take a certain value to resolve the horizon and flatness
problems of standard hot big bang cosmology. This requires
\cite{plcp} that
\begin{equation}  \label{Ntot}
\Ns\simeq61.5+\ln{\Vhi(\sgx)^{1/2}\over\Vhi(\sgf)^{1/4}}+
\frac{1-3w_{\rm rh}}{12(1+w_{\rm rh})}\ln \frac{\pi^2g_{\rm
rh*}\Trh^4}{30\Vhi(\sgf)}-\frac1{12}\ln g_{\rm rh*},
\end{equation}
where we assumed that HI is followed, in turn, by a phase of
damped inflaton oscillations with mean equation-of-state parameter
$w_{\rm rh}$, a radiation dominated era, and a matter dominated
period. Here, $\Trh$ is the reheat temperature after HI, $g_{\rm
rh*}$ is the energy-density effective number of degrees of freedom
at $\Trh$ -- for the MSSM spectrum, we take $g_{\rm rh*}=228.75$
--, $\sgx~[\se_\star]$ is the value of $\sg~[\se]$ when $k_\star$
crosses outside the inflationary horizon, and $\sgf~[\se_{\rm f}]$
is the value of $\sg~[\se]$ at the end of HI, which can be found,
in the slow-roll approximation, from the condition
\beqs\beq {\sf max}\{\widehat\epsilon(\sg_{\rm
f}),|\widehat\eta(\sg_{\rm f})|\}=1\label{srcon}\eeq
with the slow-roll parameters calculated as follows:
\beq \label{sr}\widehat\epsilon= {1\over2}\left(\frac{\Ve_{\rm
HI,\se}}{\Ve_{\rm HI}}\right)^2={1\over2J^2}\left(\frac{\Ve_{\rm
HI,\sg}}{\Ve_{\rm HI}}\right)^2
\>\>\>\mbox{and}\>\>\>\>\>\widehat\eta=\frac{\Ve_{\rm
HI,\se\se}}{\Ve_{\rm HI}}={1\over J^2}\left(\frac{\Ve_{\rm
HI,\sg\sg}}{\Ve_{\rm HI}}-\frac{\Ve_{\rm HI,\sg}}{\Ve_{\rm
HI}}{J_{,\sg}\over J}\right)\cdot \eeq\eeqs
Given that, for a power-law potential $\phi^n$, we have
\cite{turner,review} $w_{\rm rh}=(n-2)/(n+2)$, we take for our
numerics $w_{\rm rh}=1/3$, which corresponds precisely to $n=4$.
Although we expect that, in the our cases, $w_{\rm rh}$ will
deviate slightly from this value, we consider this value quite
reliable since, for low values of $\phi$, our inflationary
potentials can be well approximated by a quartic potential. As a
consequence, our set-up is largely independent from $\Trh$ -- see
\Eref{Ntot}.

\paragraph{4.1.2.} The amplitude $\As$ of the power spectrum of
the curvature perturbation generated by $\sg$ at the pivot scale
$k_\star$ must be consistent with the data \cite{plcp}:
\begin{equation}  \label{Prob}
A_{\rm s}^{1/2}=\: \frac{1}{2\sqrt{3}\, \pi} \; \frac{\Ve_{\rm
HI}(\sex)^{3/2}}{|\Ve_{\rm
HI,\se}(\sex)|}=\frac{|J(\sg_\star)|}{2\sqrt{3}\, \pi} \;
\frac{\Ve_{\rm HI}(\sg_\star)^{3/2}}{|\Ve_{\rm
HI,\sg}(\sg_\star)|}\simeq4.627\times 10^{-5},
\end{equation}
where we assume that no other contributions to the observed
curvature perturbation exist.

\paragraph{4.1.3.} The remaining inflationary observables, i.e. the
scalar spectral index $\ns$, its running $\as$, and the tensor-to-scalar
ratio $r$, which are given by
\beq\label{ns} \mbox{\ftn\sf (a)}\>\>\> \ns=\:
1-6\widehat\epsilon_\star\ +\
2\widehat\eta_\star,\>\>\>\mbox{\ftn\sf (b)}\>\>\>
\as=\:\frac23\left(4\widehat\eta_\star^2-(\ns-1)^2\right)-
2\widehat\xi_\star,\>\>\>\mbox{and}\>\>\>\mbox{\ftn\sf
(c)}\>\>\>r=16\widehat\epsilon_\star, \eeq
where
\beq
\widehat\xi=\frac{\Ve_{\rm HI,\widehat\sg} \Ve_{\rm
HI,\widehat\sg\widehat\sg\widehat\sg}}{\Ve_{\rm HI}^2}=
\frac{\Ve_{\rm HI,\sg}\,\widehat\eta_{,\sg}}{\Ve_{\rm
HI}\,J^2}+2\widehat\eta\widehat\epsilon
\eeq
and the variables with subscript $\star$ are evaluated at
$\sg=\sg_\star$, must be in agreement with the fitting of the data
\cite{plcp} with the $\Lambda$CDM$+r$ model, i.e.
\begin{equation}  \label{nswmap}
\mbox{\ftn\sf
(a)}\>\>\ns=0.968\pm0.009~~~\mbox{and}~~~\mbox{\ftn\sf
(b)}\>\>r\leq0.12,
\end{equation}
at 95$\%$ c.l. with $|\as|\ll0.01$. Although compatible with
\sEref{nswmap}{b}, the present combined \plk\ and \bcp\ results
\cite{gws} seem to favor models with values of $r$ of order
$0.01$ -- see \Eref{rgw}.

\paragraph{4.1.4.}  To avoid corrections from quantum
gravity and any destabilization of our inflationary scenario due
to higher order non-renormalizable terms, we
impose two additional theoretical constraints on our models --
keeping in mind that $\Vhi(\sg_{\rm f})\leq\Vhi(\sg_\star)$:
\beq \label{subP}\mbox{\ftn\sf (a)}\>\> \Vhi(\sgx)^{1/4}\leq1
\>\>\>\mbox{and}\>\>\>\mbox{\ftn\sf (b)}\>\> \sgx\leq1.\eeq
As we will show in \Sref{fhi3}, the UV cutoff of our model is
equal to unity (in units of $\mP$) for $\rs\leq1$ and so no
concerns regarding the validity of the effective theory arise.

\paragraph{4.1.5} The $U(1)_{B-L}$ gauge symmetry does not generate
any extra contribution to the renormalization group running of the
MSSM gauge coupling constants and so the scale $M$ and the relevant
gauge coupling constant $g_{B-L}$ can be much lower than the
values dictated by the unification of the gauge coupling constants
within the MSSM. However, for definiteness, we consider here the
most predictive case in which $g_{B-L}=g$ ($g\simeq0.7$ is the
SUSY GUT gauge coupling constant) and $M$ is determined by requiring
that $\vev{\Phi}$ and $\vev{\bar\Phi}$ take values compatible
with the unification of the MSSM gauge coupling constants. In
particular, the SUSY GUT scale
$\Mgut\simeq (2/2.433)\times10^{-2}$ is to be identified with the
lowest mass scale of the model at the SUSY vacuum in \Eref{vevs},
i.e.
\beq \label{Mpqf} {\sqrt{\cm(1-N\rs)}gM\over
\sqrt{\fr(\phi=M)}}=\Mgut\>\Rightarrow\>M=\frac{\Mgut}
{\sqrt{g^2\cm(1-N\rs)
-\cp\Mgut^2}}\simeq\frac{\Mgut}{g\sqrt{\cm}}.\eeq
The requirement that the expression $g^2\cm(1-N\rs)-
\cp\Mgut^2$ is positive sets an upper bound on $\cp$ for every
$\cm$. Namely, we should have $\cp\leq g^2(1-N\rs)\cm /\Mgut^2$,
which, however, is too loose to restrict the parameters.

\subsection{Analytic Results}\label{res}

Our analytic results are based on the tree-level inflationary
potential in \Eref{2Vhio} and are identical for both $K=\Ka$ and
$K=\Kb$ provided that $\Na$ and $\Nb$ are related to $n$ as shown
in this equation. Note that, not only the form of $\Vhio$ in
\Eref{2Vhio}, but also the canonical normalization of $\phi$ is
practically identical in the two cases -- see Eqs.~(\ref{mK}) and
(\ref{VJe}). The slow-roll parameters read
\beq \label{srg}\eph=
\frac{8(1-n\cp\sg^2)^2}{\cm\sg^2\fr}\>\>\>\mbox{and}\>\>\>
\ith=4\frac{3+\cp\sg^2\lf
n\lf4n\cp\sg^2-9\rg-2\rg}{\cm\sg^2\fr}\cdot \eeq
The termination of HI is triggered by the violation of the $\ith$
criterion at a value of $\sg$ equal to $\sgf$. Since $\sgf\ll\sgx$,
the slow-roll parameters in \Eref{srg} can be well approximated
by performing an expansion for small values of $\sg$. We find
\beq \eph\simeq8\frac{1-(1+2n)\cp\sg^2}{\cm\sg^2}
~~~\mbox{and}~~~\ith\simeq 4\frac{3-(5+9n)\cp\sg^2}{\cm\sg^2}
\,\cdot\label{sra}\eeq
Employing these expressions, $\sgf$ is calculated to be
\beqs\beq \what\eta\lf\sgf\rg=1\>\Rightarrow\>\sgf \simeq
2\sqrt{3}\,\Big(\cm+20\cp+36n\cp\Big)^{-1/2}\,. \label{sgap1}\eeq
Note that the violation of the $\what\epsilon$ criterion occurs at
$\sg=\tilde\sg_{f}$ such that
\beq \what\epsilon(\tilde\sg_{\rm f})=1\>\Rightarrow\>
\tilde\sgf\simeq2\sqrt{2}\,\Big(\cm+8\cp+16n\cp\Big)^{-1/2}<\sgf\,.
\label{sgap}\eeq\eeqs
We proceed with our analysis presenting separately our results
for the two radically different cases: the case $n=0$ in \Sref{num0}
and the case $n<0$ in \Sref{num1}.

\subsubsection{The $n=0$ Case}\label{num0}

Given that $\sgf\ll\sgx$, $\Ns$ can be calculated via \Eref{Nhi}
as follows:
\beqs\beq \label{s*}
\Ns\simeq\cm\lf{\sgx^2-\sgf^2}\rg/8\>\Rightarrow\>
\sgx\simeq2\sqrt{2\Ns/\cm}.\eeq
Obviously, HI with \sub\ values of $\sg$ can be attained if
\beq \label{fsub} \sgx\leq1~~\Rightarrow~~\cm\geq8\Ns\simeq480
\eeq\eeqs
for $\Ne_\star\simeq60$. Therefore, large values of $\cm$ are
dictated, whereas $\cp$ remains totally unconstrained by this
requirement. Replacing $\Vhio$ from \Eref{2Vhio} in \Eref{Prob},
we find
\begin{equation} \As^{1/2}=\frac{\ld\sgx^3}{32\pi}\sqrt{\frac{\cm}{3 +
3\cp \sgx^2}}
\>\>\Rightarrow\>\>\ld\simeq\pi\cm\sqrt{\frac{6\As\fns}{\Ns^{3}}}\,,
\label{lang} \eeq
where $\fns=\fr(\sgx)=1+8\rs\Ns$ (for $n=0$) and \Eref{s*} was use
in the last step. Inserting, finally, this equation into \Eref{ns},
we find the following expressions for $\ns, \as$, and $r$:
\beq \label{gns}  \ns\simeq1-\frac2\Ns\ +\ \frac1{\Ns\fns},~~
\as\simeq\frac{1-2\fns-2\fns^2}{\Ns\fns},~~~\mbox{and}~~~
r\simeq\frac{16}{\Ns\fns}\,\cdot\eeq
Therefore, a clear dependence of the observables  on $\rs$ arises.
It is worth noticing that these results coincide with the ones
obtained for the model of kinetically modified non-minimal
inflation established in \cref{nMkin} with $m=0$ and $n=4$ -- in
the notation of this reference.

\subsubsection{The $n<0$ Case}\label{num1}

When $n<0$, $\Ns$ can be estimated again through \Eref{Nhi} with
the result
\beq \Ns \simeq
\frac1{8n\rs}\ln\frac{1-n\cp\sgf^2}{1-n\cp\sgx^2}\,,
\label{Ngm}\eeq
which, obviously, cannot be reduced to the one found for $n=0$ --
cf. \Eref{s*}. Neglecting $\sgf$ in the last equality -- since
$\sgf\ll\sgx$ -- and solving w.r.t. $\sgx$, we find
\beqs\beq\label{sm*}\sgx\simeq
\sqrt{\frac{1-\re}{n\cp}},\>\>\>\mbox{where}\>\>\>\re =
e^{-8n\rs\Ns}\,.\eeq
Note that the dependence of $\sgx$ on $\Ns$ is radically different
from the one in \Eref{s*} and resembles the dependence found in
Refs.~\cite{nIG,nMCI} for $n<0$. However, $\sgx$ can again fulfill
\sEref{subP}{b} since
\beq \label{fmsub}
\sgx\leq1\>\>\>\Rightarrow\>\>\>\cm\geq\frac{1-\re}{|n|\rs},\eeq
\eeqs
where the lower bound on $\cm$ turns out to be $\rs$-dependent --
in contrast to the case of \Eref{fsub}. Substituting \Eref{sm*} into
\Eref{Prob} and solving w.r.t. $\ld$, we end up with
\begin{equation} \label{langm}
\ld\simeq32\pi\sqrt{3\As}\re\fnns^{n+1/2}\cm{\lf n\rs/(1-\re)
\rg^{3/2}},\eeq
where $\fnns=\fr(\sgx)=(1+n-\re)/n$ for $n<0$. We remark that
$\ld$ remains proportional to $\cm$ (for fixed $n$ and $\rs$) as
in the case with $n=0$ -- cf. \Eref{lang}. Inserting \Eref{sm*}
into \Eref{srg} and employing \sEref{ns}{a}, we find
\beqs\beq \label{nsgm} \ns\simeq
1+\frac{8\rs\lf2n\re^2+2(1+n)-(2+n)\re\rg}{(1-\re)\fnns}\cdot \eeq
From this expression, we see that $n<0$ and $\rs<1$ assist us to
reduce $\ns$ so as to become considerably lower than unity as
required by \sEref{nswmap}{a}. Using  Eqs.~(\ref{sm*}),
(\ref{srg}), and (\ref{ns}{\sf\ftn b, c}), we arrive at
\beq \label{ragm}
\as=\frac{64\re\rs^2\lf\re^2(2+n)(2n-1)-(1+n)(2+n)-4\re(n^2-1)
\rg}{(1-\re)^2\fnns^2}~~~\mbox{and}~~~r\simeq
\frac{128n\re^2\rs}{(1-\re)\fnns}\,\cdot \eeq\eeqs
From the last result, we conclude that mainly the fact that
$|n|\neq0$ and secondarily the fact that $n<0$ help us to
increase $r$.

\subsection{Numerical Results}\label{num}

Adopting the definition of $n$ in \Eref{2Vhio}, our models,
which are based on $W$ in \Eref{Win} and $K$ in \Eref{K1} or
(\ref{K2}), can be universally described by the following
parameters:
$$\ld,\>n,\>\cm,\>\cp,\>\kx,\>k_-,\>\mbox{and}\>k_{S-}.$$
Recall that $\Ns$ turns out to be independent of $\Trh$, as
explained in \Sref{fhi2}, and $M$, which is determined by
\Eref{Mpqf}, does not affect the inflationary dynamics and the
predictions since $M\ll\phi$ during inflation. From the remaining
parameters, $\kx$ influences only $\what m_{s}^2$ in \Tref{tab1}
and $k_{S-}$ enters only into the higher order terms -- not shown
in the formulas of \Tref{tab1} -- in the expansions of $\what
m_{\th+}^2$ and $\what m_{\th_\Phi}^2$. On the other hand, $k_-$
does not appear at all in our results. Given that the contribution
of $\dV$ to $\Vhi$ in \Eref{Vrc} can be easily tamed with a
suitable selection of $\Ld$ -- see \Tref{tab3} below --, our
inflationary outputs are essentially independent of these three
parameters, provided that we choose them so that the relevant
masses squared are positive. To ensure this, we set $\kx=k_{S-}=1$
throughout our calculation. Moreover, the bulk of our results are
independent of the choice between $K=\Ka$ or $K=\Kb$, especially
for $\rs\leq0.1$ since $J$ in \Eref{VJe} remains
undistinguishable. However, for definiteness, we present our
results for $K=\Ka$, unless otherwise stated.

For fixed $n$, the remaining three free parameters of our models
during HI, which are $\cm$, $\cp$, and $\ld$, can be reduced by one
leaving us with the two free parameters $\rs$ and $\ld/\cm$. This
fact can be understood by the following observation: If we perform
the rescalings
\beq\label{resc}
\Phi\to\Phi/\sqrt{\cm},~~\bar\Phi\to\bar\Phi/\sqrt{\cm},~~~
\mbox{and}~~~S\to S,\eeq
the superpotential $W$ in \Eref{Win} depends on $\ld/\cm$ ($M$ is
not important as we explained) and the K\"{a}hler potential $K$
in \Eref{K1} or (\ref{K2}), for $S$=0, $\Phi=\bar\Phi^*$, and fixed
$n$, depends on $\rs$. As a consequence, $\Vhio$ depends exclusively
on $\ld/\cm$ and $\rs$ via $\fr$ in \Eref{minK}. The confrontation
of these parameters with observations is implemented as follows:
Substituting $\Vhi$ from Eq.~(\ref{Vhic}) in Eqs.~(\ref{Nhi}),
(\ref{sr}), and (\ref{Prob}), we extract the inflationary observables
as functions of $n, \rs$, $\ld/\cm$, and $\sgx$. The two latter
parameters can be determined by enforcing the fulfillment of
Eqs.~(\ref{Ntot}) and (\ref{Prob}), whereas $n$ and $\rs$ largely
affect the predictions for $\ns$ and $r$ and are constrained by
Eq.~(\ref{nswmap}). Moreover, \sEref{subP}{b} bounds $\cm$ from
below, as seen from \eqs{fsub}{fmsub}. Finally, \Eref{pos} provides
an upper bound on $\rs$, which is different for $N=\Na$ and $N=\Nb$,
discriminating slightly the two cases.

We start the presentation of our results by checking the impact
of $\dV$ in \Eref{Vrc} on our inflationary predictions. This is
illustrated in \Tref{tab3}, where we arrange input and output
parameters of our model with $K=\Ka$ which are consistent with
the requirements of \Sref{fhi2}. Namely, we fix $\rs=0.015$ and
$n=-1/50$ or $n=-1/20$, which are representative values as seen
from \Fref{fig1} below. In the second and third columns of this
table, we accumulate the predictions of the model with the RCs
switched off, whereas, in the next columns, we include $\dV$.
Following the strategy adopted in \cref{Qenq}, we
determine $\Ld$ by requiring $\dV(\sgx)=0$ or $\dV(\sgf)=0$. We
can easily deduce that our results do not change after including
the RCs with either determination of $\Ld$, since $\dV$ remains
well suppressed in both cases. Note that the resulting $\Ld$ is
well below unity in the two cases with its value in the case with
$\dV(\sgx)=0$ being larger. Therefore, our findings can be
accurately reproduced by using $\Vhio$ instead of $\Vhi$. This
behavior persists even if we take $K=\Kb$. In this case, $\dV$
assumes even lower values, especially if we select $\Ld$ such
that $\dV(\sgx)=0$. This is due to the fact that the values of
$\what m_i$ entering into \Eref{Vrc} for $K=\Kb$ are different
from those for $K=\Ka$. However, this does not cause any
differentiation between the two models.
\begin{table}[!t]
\begin{center}
\renewcommand{\arraystretch}{1.3}
{\small \begin{tabular}{|l||cc||cc||cc|} \hline
\multicolumn{7}{|c|}{\sc Input Parameters}\\\hline\hline
$-1/n$ &$50$&$20$&$50$&$20$&$50$&$20$\\
$\cm/10^{2}$ &$5.03$&$22.3$&$5.03$&$22.3$&$5.03$&$22.3$\\
$\rs$&\multicolumn{2}{|c||}{$0.015$}&\multicolumn{2}{|c||}
{$0.015$}&\multicolumn{2}{|c|}{$0.015$}\\
$\sgx$ &$1$&$0.5$&$1$&$0.5$&$1$&$0.5$\\
\hline\hline
\multicolumn{7}{|c|}{\sc Output Parameters}\\\hline
&\multicolumn{2}{|c||}{$\dV=0$}&\multicolumn{2}{|c||}
{$\dV(\sgx)=0$}&\multicolumn{2}{|c|}{$\dV(\sgf)=0$}\\
\hline\hline
$\ld/10^{-3}$ &$1.13$&$5.21$&$1.13$&$5.21$&$1.13$&$5.21$\\
$\sgf/0.1$ &$1.4$&$0.67$&$1.4$&$0.67$&$1.4$&$0.67$\\
$\Ld/10^{-5}$ &$-$&$-$&$20.1$&$23.5$&$1.02$&$1.07$\\
$\Ns$ &$58.5$&$58.1$&$58.5$&$58.1$&$58.5$&$58.1$\\
\hline
$\ns/0.1$ &$9.66$&$9.68$&$9.66$&$9.68$&$9.66$&$9.68$\\
$-\as/10^{-4}$ &$6.3$&$6.4$&$6.3$&$6.4$&$6.3$&$6.4$\\
$r/0.01$&$3.8$&$4.77$&$3.8$&$4.77$&$3.8$&$4.77$\\
\hline
\end{tabular}}
\end{center}
\hfill \caption[]{\sl\small Input and output parameters of the
model which are compatible with all the requirements of \Sref{fhi2}
for $\kx=k_{S-}=1$. We use the tree-level potential by
switching off the RCs (i.e. taking $\dV=0$)
or the one-loop corrected potential with the
renormalization scale $\Ld$ determined such that $\dV(\sgx)=0$ or
$\dV(\sgf)=0$, as indicated in the table.} \label{tab3}
\end{table}

The predictions of our models can be encoded in the $\ns-\rw$
plane, where $\rw$ is the value of $r$ at the scale $k=0.002/{\rm
Mpc}$. This is shown in \Fref{fig1} for $n=0$ (solid line),
$n=-1/50$ (dashed line), and $n=-1/20$ (dot-dashed line). The
variation of $\rs$ on each line is also depicted. To obtain an
accurate comparison with the marginalized joint $68\%$ [$95\%$]
regions from the {\it Planck}, \bcp\, and BAO data -- they are
also depicted as dark [light] shaded areas --, we compute
$\rw=16\eph(\sg_{0.002})$, where $\sg_{0.002}$ is the value of
$\sg$ when the scale $k=0.002/ {\rm Mpc}$, which undergoes $\what
N_{0.002}=\Ns+3.22$ e-foldings during HI, crosses outside the
horizon of HI. For $n=0$ and $\rs\leq1$, the line is almost
straight and essentially coincides with the corresponding results
of \crefs{roest,nMkin}. For low values of $\rs$, this line
converges toward the values of $\ns$ and $\rw$ obtained within the
simplest model with a quartic potential, whereas, for larger
values of $\rs$, it crosses the observationally allowed corridors
approaching its universal attractor value \cite{roest} for
$\rs\gg1$. We cut this line at $\rs\simeq1/3$ [$\rs\simeq1/2$] for
$K=\Ka$ [$K=\Kb$], where the bound in \Eref{pos} is saturated.
This bound overshadows the one derived from the unitarity
constraint -- see \Sref{fhi3} -- and restricts $\rw$  to be larger
than $0.0028$ [$0.0019$] for $K=\Ka$ [$K=\Kb$]. For quite small
values of $\rs$, the curves corresponding to $n<0$ converge to the
curve for $n=0$. However, for larger values of $\rs$, these curves
move away from the $n=0$ line turning to the right and spanning
the observationally allowed ranges with quite natural values of
$\rs$, consistently with \eqs{nsgm}{ragm}, which are in excellent
agreement with the numerical results. Similarly to the $n=0$ case,
the $n<0$ cases too provide us with a lower bound on $r$.
Specifically, for $n=-1/50$ [$n=-1/20$], we obtain $\rw\geq0.0123$
[$\rw\geq0.03$]. Finally, we remark that, for any $n$, we can
define a minimal $r_\pm^{\rm min}$ and a maximal $r_\pm^{\rm max}$
value of $\rs$ in the marginalized joint $95\%$ region.
Specifically, we find
\beq\bem \label{res1} & r_\pm^{\rm min}\simeq
4.5\times10^{-3}&~~~~~\mbox{and}~~~~r_\pm^{\rm
max}\simeq1/3~[1/2]&\hspace{-.6cm}\mbox{for}~~~n=0,\cr
&r_\pm^{\rm min}\simeq4.8\times10^{-3}&\mbox{and}~~~~r_\pm^{\rm
max}\simeq0.13~~~&~~~\mbox{for}~~~ n=-1/50,\cr & r_\pm^{\rm
min}\simeq5.0\times10^{-3}&~~\mbox{and}~~~~r_\pm^{\rm
max}\simeq0.052~~~&~~~\mbox{for}~~~ n=-1/20,\eem\eeq
where the values given for $n=0$ correspond to $K=\Ka~[K=\Kb]$.
Note that increasing $|n|$, $r_\pm^{\rm max}$ decreases and
becomes more natural according to the argument below \Eref{K2}.

\begin{figure}[!t]\vspace*{-.25in}
\begin{center}
\epsfig{file=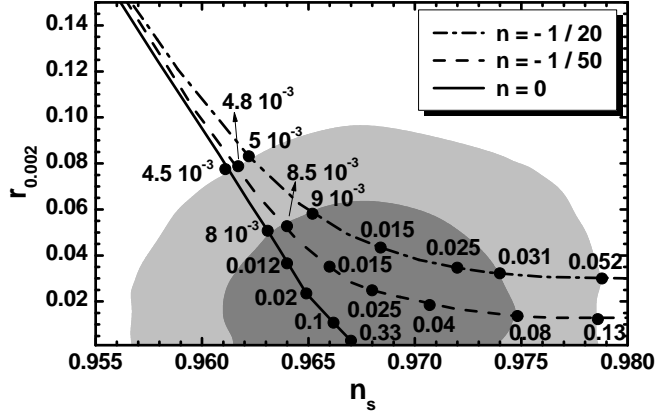,height=3.65in,angle=-90}\ec
\vspace*{-.2in}\hfill \caption{\sl \small Predicted curves in the
$\ns-\rw$ plane for  $n=0$ (solid line), $n=-1/50$ (dashed line),
$n=-1/20$ (dot-dashed line), $\kx=k_{S-}=1$, and various values of $\rs$
indicated on the curves. The marginalized joint $68\%$ [$95\%$]
regions from the Planck, \bcp\, and BAO data are depicted as dark
[light] shaded areas.}\label{fig1}
\end{figure}

The structure of $\Vhi$ as a function of $\sg$ for $\sgx=1$,
$\rs=0.015$, and the values of $n$ employed in \Fref{fig1} is
displayed in \Fref{fig2}. The values of $n, \ld$, and $\rs$, shown
in this figure, yield $\ns=0.964, 0.966$, or $0.968$ and $r=0.033,
0.038$, or $0.047$ for $n=0, -1/50$, or $-1/20$ respectively. The
corresponding values of $\cm$ are $(4.66, 5.03,{\rm or}~5.6)\times
10^2$, whereas the values of $\se_\star$ derived from \Eref{se1}
are $13.87$, $14.17$, or $14.6$. We verify that observable values
of $r$ are associated with \trns\ values of $\se_\star$ in
accordance with the Lyth bound \cite{lyth}. This fact, though,
does not invalidate our scenario since the corresponding values of
the initial inflaton $\sg$, which is directly related to the
superfields $\Phi$ and $\bar\Phi$ appearing in the definition of
our models in \eqs{Win}{K1} or (\ref{K2}), remain subplanckian --
cf. \sEref{subP}{b}. We also remark that, in all cases,
$\Vhi^{1/4}(\sgx)$ turns out to be close to the SUSY GUT scale
$M_{\rm GUT}\simeq 8.2\times10^{-3}$, which is imperative -- see
e.g. \cref{rRiotto} -- for achieving values of $r$ close to $0.1$.
We finally observe that the slope of $\Vhi$ close to $\sg=\sgx$
increases with $|n|$. This is expected to elevate $\eph$ -- see
\Sref{res} -- and, via \sEref{ns}{c}, $r$ too.

\begin{figure}[!t]\vspace*{-.25in}
\begin{center}
\epsfig{file=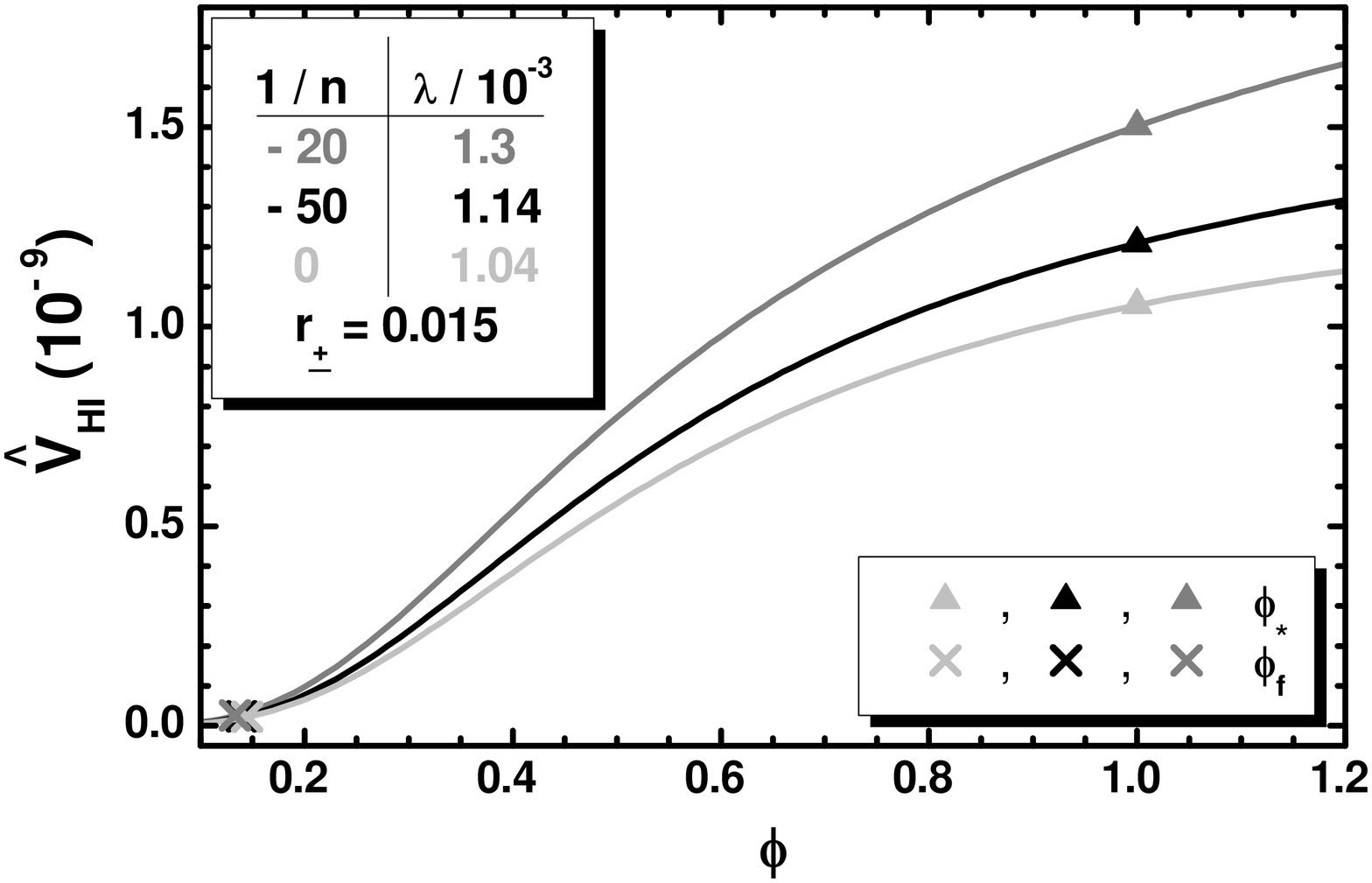,height=3.65in,angle=-90}\ec
\vspace*{-.2in}\hfill \hfill \caption[]{\sl \small The inflationary
potential $\Vhi$ as a function of $\sg$ for $\sg>0$,
$\rs\simeq0.015$, and $n=0$, $\ld=1.04\times 10^{-3}$ (light gray line),
$n=-1/50$, $\ld=1.14\times 10^{-3}$ (black line), or $n=-1/20$,
$\ld=1.3\times 10^{-3}$ (gray line). The values of $\sgx$ and $\sgf$ are
also indicated.}\label{fig2}
\end{figure}

\begin{figure}[!t]
\hspace*{-.1in}
\begin{minipage}{8in}
\epsfig{file=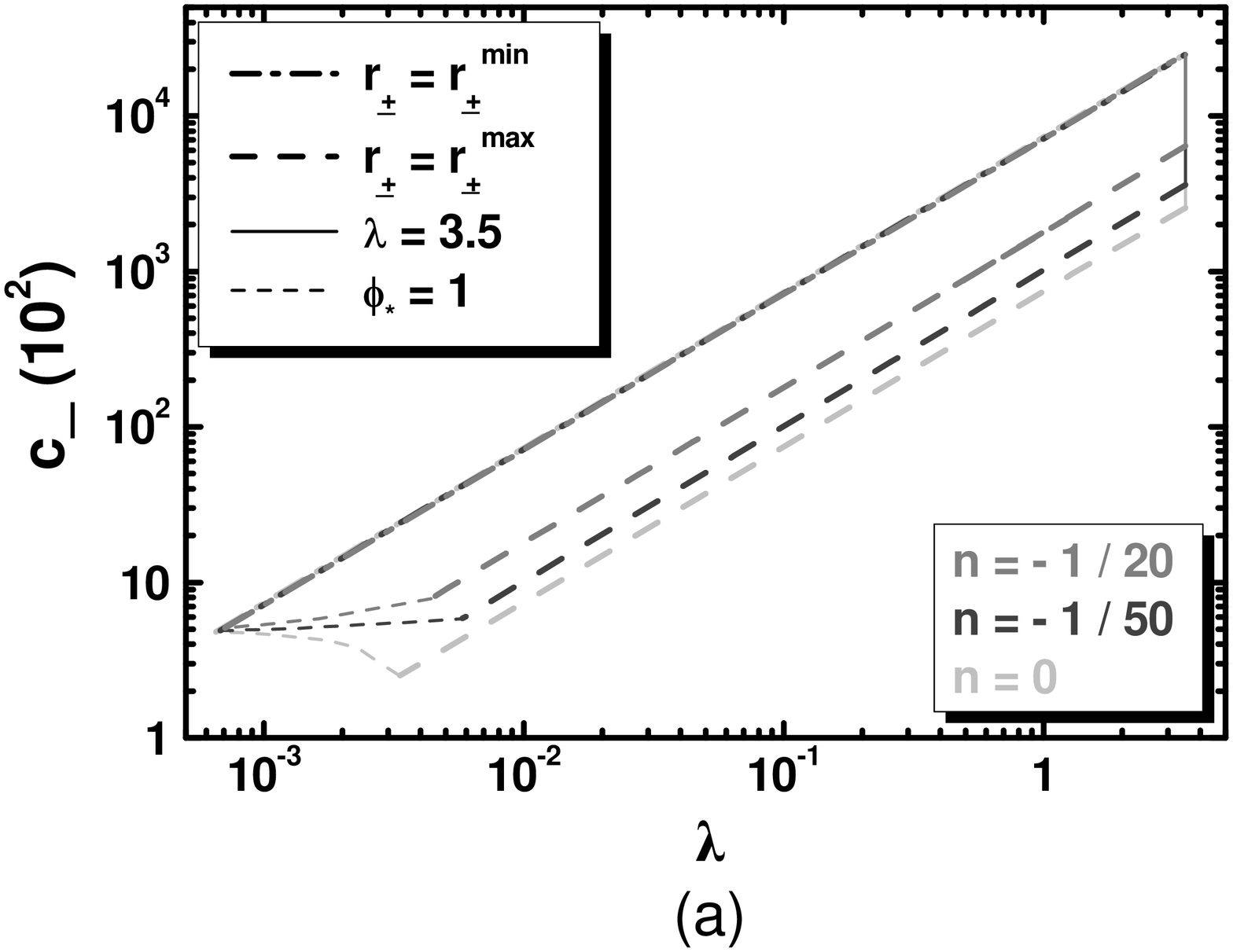,height=3.5in,angle=-90}
\hspace*{-1.cm}
\epsfig{file=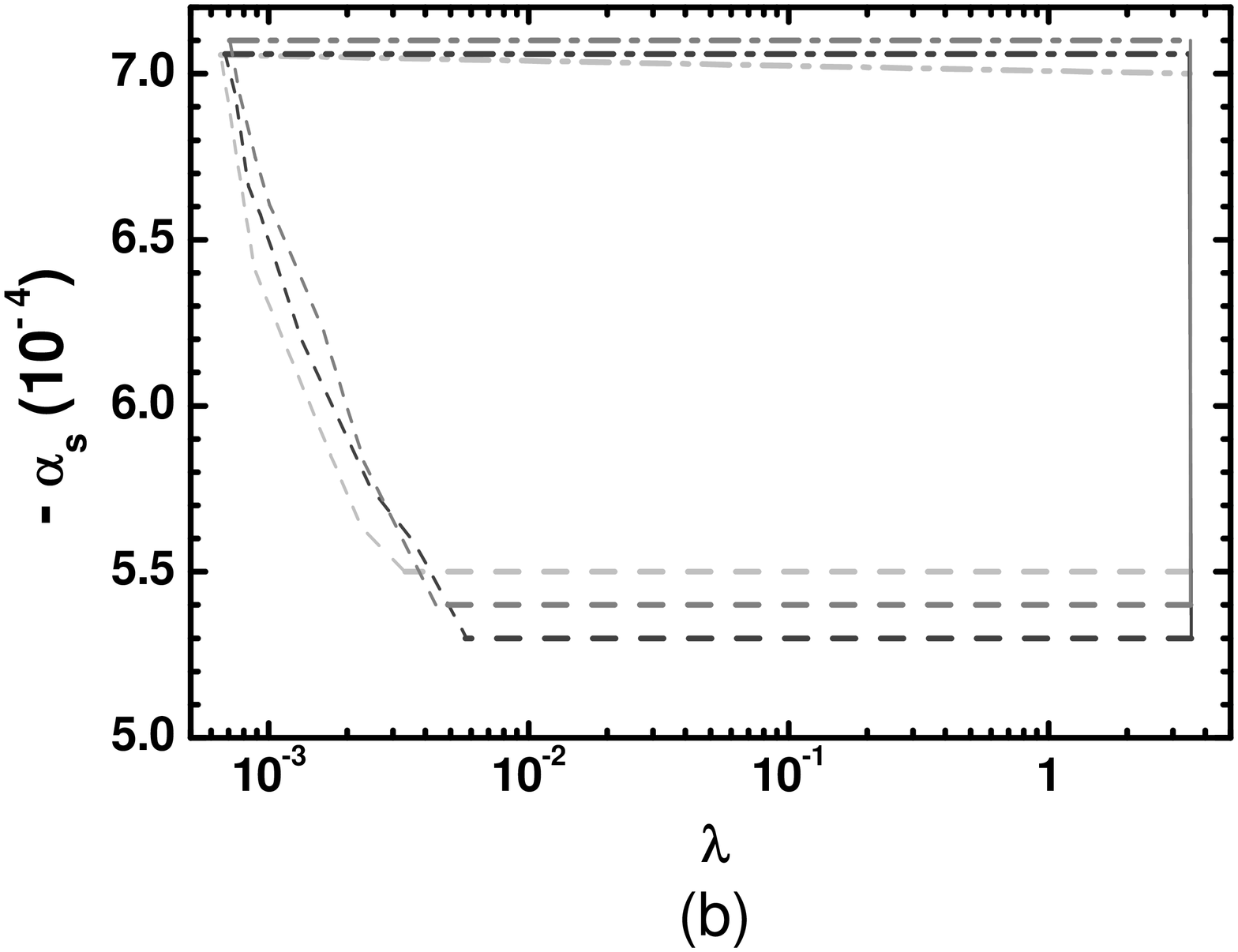,height=3.5in,angle=-90} \hfill
\end{minipage}
\vspace*{-.0in} \hfill \caption[]{\sl\small Allowed regions in the
$\ld-\cm$ ({\sffamily\ftn a}) and $\ld-\as$ ({\sffamily\ftn b})
plane for $\kx=k_{S-}=1$ and $n=0, -1/50$, or $-1/20$ bounded by
light, dark, or normal gray lines, respectively. The conventions
adopted for the various lines are shown in the panel (a).}
\label{fig3}
\end{figure}

To specify further the allowed ranges of the parameters of our
models, we plot in \sFsref{fig3}{a} and -{\sf\ftn (b)} the allowed
regions in the $\ld-\cm$ and $\ld-\as$ planes, respectively. The
conventions adopted for the various lines are shown in
\sFref{fig3}{a}. In particular, the boundary curves of the allowed
region for $n=0, -1/50$, or $-1/20$ are represented by light, dark,
and normal gray lines, respectively. The dot-dashed and dashed lines
correspond, respectively, to the minimal and maximal values of $\rs$
given in \Eref{res1}, whereas, on the thin short-dashed lines, the
constraint of \sEref{subP}{b} is saturated. The perturbative bound
$\ld\leq 3.5$ (so that the expansion parameter $\ld^2/4\pi\leq 1$)
limits the various regions at the opposite end by a thin
solid line. For $K=\Kb$, the light gray dashed line is expected to
be transported to lower values of $\cm$, but keeping the same slope.
Note that the dot-dashed lines coincide with each other in
\sFref{fig3}{a} and this is almost the case with the dark and light
gray dot-dashed lines in \sFref{fig3}{b} too. We observe that
$\ld/\cm$ remains constant for fixed $\rs$, as expected by the
argument below \Eref{resc}. The required $\cm$ for \sub\ excursions
of $\phi$ -- overall, we obtain $0.014\leq\sg\leq1$ -- is quite
large and increases with $|n|$. On the other hand, $|\as|$ remains
sufficiently low. Therefore, our models are consistent with the
fitting of the data with the $\Lambda$CDM+$r$ model \cite{plcp}.

\begin{figure}[!t]\vspace*{-.in}
\begin{center}
\epsfig{file=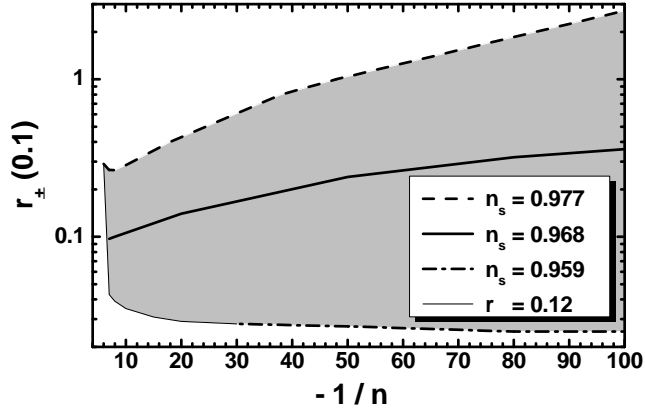,height=3.65in,angle=-90}\ec
\vspace*{-.2in}\hfill \caption[]{\sl \small Allowed (shaded)
region compatible with Eqs.~(\ref{Ntot}), (\ref{Prob}),
(\ref{nswmap}), and (\ref{subP}) in the $(-1/n)-\rs$ plane for
$\kx=k_{S-}=1$. The conventions adopted for the various lines are
also shown.}\label{fig4}
\end{figure}

Concentrating on the most promising cases with $n<0$, we delineate,
in \Fref{fig4}, the overall allowed region of our models by
varying continuously $n$ and $\rs$. The conventions adopted for
the various lines are also shown in the figure. In particular, the
dashed [dot-dashed] line corresponds to $\ns=0.977$ [$\ns=0.959$],
whereas the solid line is obtained by fixing $\ns=0.968$ -- see
\sEref{nswmap}{a}. On the thin solid boundary line, the bound in
\sEref{nswmap}{b} is saturated. We remark that, as $\rs$ increases
with fixed $n$, $\ns$ increases too, while $r$ decreases. This
agrees with our findings in \Fref{fig1}. Note that, for
$n\geq-1/30$, the thick dot-dashed and the thin solid lines
coincide. Overall, for $\ns=0.968$ and $\Ns\simeq58$, we find:
\beq\label{resgmr} 1.86\lesssim 10^6{\ld\over
\cm}\lesssim3.6\>\>\>\mbox{with}\>\>\>8\lesssim
-\frac1n\lesssim100,\>\>\> 1\lesssim
{\rs\over0.01}\lesssim3.6\,,\>\>\>\mbox{and}\>\>\> 1.8\lesssim
{r\over0.01}\lesssim12\,. \eeq
From these results, we infer that $\rs$ takes more natural --
in the sense of the discussion below \Eref{K2} -- values (lower
than unity) for larger values of $|n|$.

\section{The Effective Cut-off Scale}\label{fhi3}

An outstanding trademark of our setting is that perturbative
unitarity is retained up to $\mP$, despite the fact that the
implementation of HI with \sub\ values of $\phi$ requires
relatively large values of $\cm$ -- see \eqs{fsub}{fmsub}. To show
this, we extract the UV cut-off scale $\Qef$ of the effective
theory following the systematic approach of \cref{riotto}. Let us
first clarify that, although the expansions presented below about
$\vev{\phi}=M\ll1$ are not valid \cite{cutof} during HI, we
consider the resulting $\Qef$ as the overall cut-off scale of the
theory, since reheating is regarded as an unavoidable stage of the
inflationary dynamics \cite{riotto}.

The canonically normalized inflaton can be written as follows -- see
\Eref{VJe}:
\beq\dphi=\vev{J}\dph\>\>\>\mbox{with}\>\>\>\dph\equiv \phi-M
\>\>\>\mbox{and}\>\>\>\vev{J}=
\sqrt{\vev{\kp_+}}\simeq\sqrt{\cm/\vev{\fr}}\,,\label{dphi}
\eeq
where the last (approximate) equality is valid only for $\rs\ll1/N$
-- see \Eref{pos}. Note, in passing, that the mass of $\dphi$ at
the vacuum is calculated to be
\beq \label{masses} \msn=\left\langle\Ve_{\rm
HI0,\se\se}\right\rangle^{1/2}= \left\langle \Ve_{\rm
HI0,\sg\sg}/J^2\right\rangle^{1/2}\simeq\frac{\ld
M}{\sqrt{2\cm(1-N\rs)}}\eeq
with numerical values $(1.5-3.3)\times10^{-8}$ along the
solid line of \Fref{fig4} where $\ns\simeq0.968$. Given that
$\vev{\fr}\simeq1$, $|n|\ll1$, and $\ld/\cm$ is fixed -- see
\eqs{lang}{langm} --, the variation of $\msn$ is mainly generated
by the variation of $\rs$. In other words, for fixed $n$, $\msn$
depends only on $\rs$ and not on $\ld, M, \cm$, or $\cp$
separately. Note that the resulting values of $\msn$ are almost
two orders of magnitude lower than the values obtained in similar
models \cite{nIG, nmH, nMCI} and so the successful activation of
the mechanism of non-thermal leptogenesis \cite{ntlepto} is an
important open issue.

The fact that $\dphi$ does not coincide with $\dph$ in \Eref{dphi}
-- contrary to the case of standard non-minimal HI \cite{cutoff,cutof}
-- ensures that our models are valid up to $\mP=1$ -- cf. \cref{nMkin}.
Taking into account \Eref{Snik} and calculating the action ${\sf S}$ in
\Eref{Saction1} on the path of \Eref{inftr} with spatially constant
values of $\phi$, we find
\beqs \beq\label{S3} {\sf S}=\int d^4x \sqrt{-\what{
\mathfrak{g}}}\lf-\frac{1}{2} \rce +\frac12\,J^2
\dot\phi^2-\Ve_{\rm HI0}+\cdots\rg, \eeq
where $J$ is given in \Eref{VJe} and $\Vhio$ in
\Eref{2Vhio}. Expanding $J^2$ about $\phi=\vev{\phi}=M\simeq 0$
and expressing the result in terms of $\se$ ($\simeq\dphi$) using
\Eref{dphi}, we obtain
\beq J^2 \dot\phi^2=\lf1-\rs\what{\sg}^2+3N\rs^2\what{\sg}^2+
\rs^2\what{\sg}^4-5N\rs^3\what{\sg}^4+\cdots\rg\dot\se^2.
\label{Jexp}
\eeq
If we insert \Eref{cR} into \eqs{mK}{VJe} and repeat the same
expansion, we get the notorious term $2N \ca^2\se^2$
\cite{cutoff,cutof}, which entails $\Qef\ll1$ for large $\ca$. In
our case, however, $\ca$ is replaced by $\rs$ which remains low
enough -- although $\cp$ and $\cm$ may be large -- rendering the
effective theory unitarity safe. Expanding similarly $\Vhio$ in
terms of $\se$, we have
\beq\label{Vexp}
\Vhio=\frac{\ld^2\what{\sg}^4}{16\cm^{2}}\lf1-2(1+n)\rs
\what{\sg}^{2}+(3+5n+2n^2)\rs^2\what{\sg}^4-\cdots\rg\cdot
\eeq\eeqs
This expression, for $n=0$, reduces to the one presented in
\cref{nMkin}, whereas the expression for $J$ in \Eref{Jexp}
differs slightly from the corresponding one in this reference, due
to the different normalization of $\sg$ in \eqs{mK}{VJe}. Since
the positivity of $\kp_-$ in \Eref{mK} entails $\rs<1/N<1$, our
overall conclusion is that our models do not face any problem with
perturbative unitarity up to $\mP$.

\section{Conclusions}\label{con}

We presented models of Higgs inflation in SUGRA, which accommodate
inflationary observables covering the `sweet' spot of the recent
joint analysis of the \bcp\ and \plk~data. Our models, at the
renormalizable level, are tied to a unique superpotential
determined by an $R$ and a gauge $U(1)_{B-L}$ symmetry. We
selected two possible \Kap s $K$, one logarithmic and one
semi-logarithmic -- see \eqs{K1}{K2} --, which respect the
symmetries above and a mildly violated shift symmetry. Both $K$'s
lead to practically identical inflationary models. The key-point
in our proposal is that the coefficient $\cm$ of the
shift-symmetric term in the \Kap s does not appear in the
supergravity inflationary scalar potential expressed in terms of
the original inflaton field, but strongly dominates the canonical
normalization of this inflaton field. The inflationary scenario
depends essentially on three free parameters ($n$, $\ld/\cm$, and
$\rs$), which are constrained to natural values leading to values
of $\ns$ and $r$ within their $1-\sigma$ observational margins.
Indeed, adjusting these parameters in the ranges $n=-(0.1-0.01)$,
$\ld/\cm=(1.86-3.6)\times10^{-6}$, and $\rs=0.01-0.036$, we obtain
$\ns\simeq0.968$ and $0.018\lesssim r\lesssim0.12$ with negligibly
small values of $|\as|$. Imposing a lower bound on $\cm$, we
succeeded to realize HI with \sub\ values of the original
inflaton, thereby stabilizing our predictions against possible
higher order corrections in the superpotential and/or the \Kap s.
Moreover, the corresponding effective theory remains trustable up
to $\mP$. We also showed that the one-loop RCs remain subdominant
for  a convenient choice of the renormalization scale. Finally,
the scale of $U(1)_{B-L}$ breaking can be identified with the SUSY
GUT scale and the mass $\msn$ of the normalized inflaton is
confined in the range $(1.5-3.3)\times10^{-8}$.

As a last remark, we would like to point out that, although we
have restricted our discussion to the $\Gbl$ gauge group, the HI
analyzed in this paper has a much wider applicability. It can be
realized within other SUSY GUTs too based on a variety of gauge
groups -- such as the left-right, the Pati-Salam, or the flipped
$SU(5)$ group -- provided that a conjugate pair of Higgs
superfields is used in order to break the symmetry. In these
cases, the inflationary predictions are expected to be quite
similar to the ones discussed here. The discussion of the
stability of the inflationary trajectory may, though, be
different, since different Higgs superfield representations may be
involved in implementing the GUT gauge symmetry breaking to
$G_{\rm SM}$.

\acknowledgments{We would like to thank I. Antoniadis, S.~Antusch,
F. Bezrukov, Wan-il Park, and M. Postma for useful discussions.
This research was supported by the MEC and FEDER (EC) grants
FPA2011-23596 and the Generalitat Valenciana under grant
PROMETEOII/2013/017.}


\newcommand\jcap[3]{{\it J. Cosmol. Astropart. Phys.} {\bf #1}~(#2)~#3 }


\begin{thebibliography}{10} {\baselineskip 2.pt \ftn

\bibitem{gws} P.A.R.~Ade {\it et al.} [\bcp\ and \plk\ Collaborations],
\prl{114}{2015}{101301} [\arxiv{15 02.00612}].

\bibitem{plcp} P.A.R.~Ade {\it et al.} [\plk\ Collaboration],
\arxiv{1502.02114}.

\bibitem{gws2} M.J.~Mortonson and U.~Seljak, \jcap{10}{2014}{035}
[\arxiv{1405.5857}].

\bibitem{kdust} C. Cheng, Q. G. Huang, and S. Wang,
\jcap{12}{2014}{044} [\arxiv{1409.7025}]; \\ L.~Xu,
\arxiv{1409.7870}.

\bibitem{gws1} P.A.R.~Ade \etal\ [\bicep\ Collaboration],
{\sl Phys. Rev. Lett.} {\bf 112} (2014) 241101
[\arxiv{1403.3985}].

\bibitem{aroest}  R.~Kallosh, A.~Linde, and D.~Roest,
\jhep{11}{2013}{198} [\arxiv{1311.0472}]; \\  R.~Kallosh,
A.~Linde, and D.~Roest, \jhep{08}{2014}{052} [\arxiv{1405.3646}].

\bibitem{rEllis} J.~Ellis, M.~Garcia, D.~Nanopoulos, and
K.~Olive, \jcap{05}{2014}{037} [\arxiv{1403.7518}]; \\
J.~Ellis, M.~Garcia, D.~Nanopoulos, and K.~Olive,
\jcap{08}{2014}{044} [\arxiv{1405.0271}].

\bibitem{nIG}  C.~Pallis, \jcap{04}{2014}{024}
[\arxiv{1312.3623}]; \\ C.~Pallis, \jcap{08}{2014}{057}
[\arxiv{1403.5486}];\\ C.~Pallis, \jcap{10}{2014}{058}
[\arxiv{1407.8522}].

\bibitem{nMCI} C. Pallis and Q. Shafi, \prd{86}{2012}{023523}
[\arxiv{1204.0252}];\\ C. Pallis and Q. Shafi,
\jcap{03}{2015}{023} [\arxiv{1412.3757}].

\bibitem{nMkin} C. Pallis, \prd{91}{2015}{123508} [\arxiv{1503.05887}].

\bibitem{shift} M.~Kawasaki, M.~Yamaguchi, and T.~Yanagida,
\prl{85}{2000}{3572} [\hepph{0004243}];\\
P.~Brax and J.~Martin,
{\it Phys. Rev.} {\bf D 72} (2005) 023518 [\hepth{0504168}];
\\S. Antusch, K. Dutta, and P.M. Kostka, \plb{677}{2009}{221}
[\arxiv{0902.2934}];\\ R.~Kallosh, A.~Linde, and T.~Rube,
\prd{83}{2011}{043507} [\arxiv{1011.5945}];\\ T.~Li, Z.~Li, and
D.V.~Nanopoulos, \jcap{02}{2014}{028} [\arxiv{1311.6770}];\\
K.~Harigaya and T.T.~Yanagida, \plb{734}{2014}{13}
[\arxiv{1403.4729}]; \\ A.~Mazumdar, T.~Noumi, and M.~Yamaguchi,
\prd{90}{2014}{043519} [\arxiv{1405.3959}];\\  C.~Pallis and
Q.~Shafi, \plb{736}{2014}{261} [\arxiv{1405.7645}].

\bibitem{bauman} D.~Baumann and D.~Green, \jhep{09}{2010}{057}
[\arxiv{1004.3801}].

\bibitem{shiftHI} I.~Ben-Dayan and M.B.~Einhorn,
\jcap{12}{2010}{002} [\arxiv{1009.2276}].

\bibitem{takahashi} K.~Nakayama and F.~Takahashi,
\jcap{11}{2010}{009} [\arxiv{1008.2956}];\\ K.~Nakayama and
F.~Takahashi, \jcap{11}{2010}{039} [\arxiv{1009.3399}].

\bibitem{khalil} S.~Antusch {\it et al.}, \jhep{08}{2010}{100}
[\arxiv{1003.3233}];\\ L.~Heurtier, S.~Khalil, and A.~Moursy,
\arxiv{1505.07366}.

\bibitem{jones2} J.L.~Cervantes-Cota and H.~Dehnen, \prd{51}{1995}{395}
[\astroph{9412032}];\\ M.~Arai, S.~Kawai, and N.~Okada, {\it Phys. Rev.}
{\bf D 84} (2011) 123515 [\arxiv{1107.4767}];\\
M.B.~Einhorn and D.R.T.~Jones, \jcap{11}{2012}{049}
[\arxiv{1207.1710}];\\  J.~Ellis, H.J.~He, and Z.Z.~Xianyu, {\it
Phys. Rev.} {\bf D 91} (2015) 021302 [\arxiv{1411.5537}];\\
J.~Ellis, T.E.~Gonzalo, J.~Harz, and W.C.~Huang,
\jcap{03}{2015}{039} [\arxiv{1412.1460}].

\bibitem{linde1} M.B.~Einhorn and D.R.T.~Jones,
\jhep{03}{2010}{026} [\arxiv{0912.2718}]; \\ H.M.~Lee,
\jcap{08}{2010}{003} [\arxiv{1005.2735}]; \\ S.~Ferrara {\it et al.},
\prd{83}{2011}{025008} [\arxiv{1008.2942}];\\ C.~Pallis and
N.~Toumbas, \jcap{02}{2011}{019} [\arxiv{1101.0325}].

\bibitem{nmH} C.~Pallis and N.~Toumbas,
\jcap{12}{2011}{002} [\arxiv{1108.1771}];\\ C.~Pallis and
N.~Toumbas, \arxiv{1207.3730}.

\bibitem{old}  D.S.~Salopek, J.R.~Bond, and J.M.~Bardeen,
{\it Phys. Rev.} {\bf D 40} (1989) 1753; \\
R.~Fakir and W.G.~Unruh, {\it Phys. Rev.} {\bf D 41} (1990) 1792.

\bibitem{sm1} J.L.~Cervantes-Cota and H.~Dehnen, \npb{442}{1995}{391}
[\astroph{9505069}];\\ F.L.~Bezrukov and M.~Shaposhnikov,
\plb{659}{2008}{703}  [\arxiv{0710.3755}];\\
A.O.~Barvinsky {\it et al.}, \jcap{11}{2008}{021} [\arxiv{0809.2104}];\\
A.~De Simone, M.P.~Hertzberg, and F.~Wilczek, \plb{678}{2009}{1}
[\arxiv{0812.4946}].

\bibitem{linde2} R.~Kallosh and A.~Linde, \jcap{11}{2010}{011}
[\arxiv{1008.3375}];\\ R.~Kallosh, A.~Linde, and A.~Westphal, {\it
Phys. Rev.} {\bf D 90} (2014) 023534 [\arxiv{1405.0270}].

\bibitem{roest} R. Kallosh, A. Linde, and D. Roest,
{\it Phys. Rev. Lett.} {\bf 112} (2014) 011303
[\arxiv{1310.3950}].



\bibitem{cw} S.R.~Coleman and E.J.~Weinberg, \prd{7}{1973}{1888}.

\bibitem{Qenq} K.~Enqvist and M.~Kar\v ciauskas, \jcap{02}{2014}{034} [\arxiv{1312.5944}].

\bibitem{cutoff} J.L.F.~Barbon and J.R.~Espinosa,
\prd{79}{2009}{081302} [\arxiv{0903.0355}];\\
C.P.~Burgess, H.M.~Lee, and M.~Trott, \jhep{07}{2010}{007}
[\arxiv{1002.2730}];\\ M.P.~Hertzberg, \jhep{11}{2010}{023}
[\arxiv{1002.2995}].

\bibitem{cutof} F.~Bezrukov, A.~Magnin, M.~Shaposhnikov, and
S.~Sibiryakov, \jhep{01}{2011}{016} [\arxiv{1008. 5157}].

\bibitem{riotto} A.~Kehagias, A.M.~Dizgah, and A.~Riotto,
\prd{89}{2014}{043527} [\arxiv{1312.1155}].

\bibitem{ntlepto} G. Lazarides and Q. Shafi, {\it Phys. Lett.} {\bf B 258}
(1991) 305; \\ K. Kumekawa, T. Moroi, and T. Yanagida, {\it Prog. Theor. Phys.}
{\bf 92} (1994) 437 [\hepph{9405337}]; \\ G. Lazarides, Q. Shafi, and
N.D.~Vlachos, \plb{427}{1998}{53} [\hepph{9706385}]; \\ G.~Lazarides and
N.D.~Vlachos, \plb{459}{1999}{482} [\hepph{9903511}]; \\ G. Lazarides,
\hepph{9905450}.

\bibitem{strings}
P.A.R. Ade {\it et al.} [\plk\ Collaboration],
{\it Astron. Astrophys.} {\bf 571} (2014) A25 [\arxiv{1303.5085}].

\bibitem{susyhybrid} G.R. Dvali, Q. Shafi, and R.K. Schaefer,
\prl{73}{1994}{1886} [\hepph{9406319}];\\
C. Pallis and Q. Shafi, {\it Phys. Lett.} {\bf B 725} (2013) 327
[\arxiv{1304.5202}]; \\ M.~Civiletti,
C.~Pallis, and Q.~Shafi, {\it Phys. Lett.} {\bf B 733} (2014) 276
[\arxiv{1402.6254}].

\bibitem{chios} T.~Dent, G.~Lazarides, and R.~Ruiz de Austri,
{\it Phys. Rev.} {\bf D 69} (2004) 075012 [\hepph{0312033}]; \\
G.~Lazarides, {\sl Lect. Notes Phys.} {\bf 720} (2007) 3 [\hepph{0601016}].

\bibitem{talk} C. Pallis, {\it PoS \bf{CORFU2012}} (2013) 061
[\arxiv{1307.7815}].

\bibitem{postma} D.P.~George, S.~Mooij, and M.~Postma,
\jcap{11}{2014}{043} [\arxiv{1207.6963}];\\ D.P.~George, S.~Mooij,
and M.~Postma, \jcap{02}{2014}{024} [\arxiv{1310.2157}].

\bibitem{review} D.H.~Lyth and A.~Riotto, {\it Phys.
Rept.} {\bf 314} (1999) 1 [\hepph{9807278}];  \\ G.~Lazarides,
{\it J. Phys. Conf. Ser.} {\bf 53} (2006) 528
[\hepph{0607032}]; \\ A.~Mazumdar and J.~Rocher, {\it
Phys. Rept.} {\bf 497} (2011) 85 [\arxiv{1001.0993}];  \\
J.~Martin, C.~Ringeval, and V.~Vennin, {\it Phys. Dark Univ.}
{\bf 5} (2014) 75 [\arxiv{1303.3787}].


\bibitem{turner} M.S.~Turner, {\it Phys. Rev.} {\bf D 28} (1983) 1243.

\bibitem{lyth} D.H.~Lyth, \prl{78}{1997}{1861}
[\hepph{9606387}]; \\ R.~Easther, W.H.~Kinney, and B.A.~Powell,
\jcap{08}{2006}{004} [\astroph{0601276}]; \\ D.H.~Lyth,
\jcap{11}{2014}{003} [\arxiv{1403.7323}].

\bibitem{rRiotto} A.~Kehagias and A.~Riotto, \prd{89}{2014}{101301}
[\arxiv{1403.4811}].

}

\end{thebibliography}
\end{document}